\documentclass[journal]{IEEEtran}
\usepackage{amsmath,amsfonts}
\usepackage{algorithmic}
\usepackage{algorithm}
\usepackage{array}
\usepackage{xcolor}
\usepackage[caption=false,font=normalsize,labelfont=sf,textfont=sf]{subfig}
\usepackage{textcomp}
\usepackage{stfloats}
\usepackage{url}
\usepackage{verbatim}
\usepackage{graphicx}
\usepackage{cite}
\usepackage{orcidlink}
\usepackage{threeparttable}  
\usepackage{booktabs}        

\begin{document}

\title{A  Reconfigurable Computing In-Memory Macro with  Charge-sharing-based Weighted Accumulator \\


\author{
    \IEEEauthorblockN{
        Junyi Yang\orcidlink{0000-0002-5867-4943},~\IEEEmembership{Graduate Student Member,~IEEE},
        Shuai Dong\orcidlink{0009-0007-4807-5094},
        Zhengnan Fu\orcidlink{0009-0009-1235-8521},
        Hongyang Shang\orcidlink{0009-0007-6276-1947},\\
       and Arindam Basu\orcidlink{0000-0003-1035-8770},~\IEEEmembership{Fellow,~IEEE}
    }

    \thanks{
        This work was sponsored in part by RGC (C7003-24Y) and Innovation technology Fund Mid-Stream Research program under Grant ITS/018/22MS. \textit{(Corresponding authors: Shuai Dong, Arindam Basu)} 
        
      Junyi Yang, Shuai Dong, Zhengnan Fu, Hongyang Shang, Arindam Basu are with Department of Electrical Engineering,
      City University of Hong Kong, Hong Kong, China. (e-mail: shuaidong6-c@my.cityu.edu.hk; arinbasu@cityu.edu.hk.).


}
    \IEEEauthorblockA{
    }
}


}

\maketitle

\begin{abstract}

SRAM-based analog computing-in-memory  demonstrates outstanding efficiency. However, it faces three critical challenges: significant ADC overhead, high latency for multi-bit inputs, and limited read bitline voltage. To address these issues, this work proposes a multi-bit highly reconfigurable 256$\times$128 in-memory computing array supporting 1-7b input, 2-4b weight, and 1-7b output. The design employs bit-slicing (BS) with a charge-sharing-based binary-weighted accumulator (BSCHA) and a reconfigurable (1-7b) in-memory analog-to-digital converter (IMADC) with shared voltage references. Three key innovations are introduced: 1) The IMADC occupies only 3\% area overhead, achieving a 9× improvement compared to previous IMADC; 2) The BSCHA reduces latency by 1.9× and 6.6× compared to traditional pulse-width modulation (PWM) and bit-slicing modes, respectively; 3) A dual-8T bitcell enabling ternary weight storage through a decoupled read path, integrated with a read wordline under-driven cascode technique, improves linearity of unit discharge current by 7× and increases the usable read bitline voltage by 3.5×. Using noise resilient training, we show software performance for a MLP on MNIST, VGG-8 on CIFAR-10, and  Vision Transformer (ViT) on CIFAR-100 with respective accuracy reductions of only 0.1\%, 0.4\% and 0.4\% due to non-idealities. The proposed macro demonstrates high energy/area efficiency  (1023.2 TOPS/W, 27 TOPS/mm\textsuperscript{2} at 1/2/1b and 8.4 TOPS/W, 0.014 TOPS/mm\textsuperscript{2} at 7/4/7b) in 65 nm CMOS. It increases throughput (by 1.9×) and linearity (by 23×) compared to input pulse-width modulation by using BSCHA. Compared to conventional BS with digital accumulation after ADC, this method has 1.5×/6.6× better normalized energy-efficiency/throughput by reducing ADC operations in macro-level. The system-level hardware performance of VGG-8 on CIFAR-10 is evaluated through combined SPICE and NeuroSim simulations, demonstrating the significant 6× enhancements in normalized energy efficiency.

\end{abstract}


\section{Introduction}
The proliferation of deep neural networks (DNNs) has revolutionized the field of artificial intelligence (AI), enabling state-of-the-art performance in tasks, including computer vision\cite{ li2023towards}, speech processing\cite{hinton2012deep}, and image classification \cite{zhang2022quantifying},\cite{ guo202428}.  However, the implementation of DNNs on conventional von Neumann computing architectures faces fundamental limitations, manifested through the memory wall phenomena\cite{guo202428, lee202328}. These bottlenecks originate from the energy-intensive data shuttling between processing units and memory hierarchies. The challenge is further compounded by the escalating scale of modern DNN models, where the storage and frequent access of millions to billions of parameters introduce prohibitive energy overheads and latency penalties that severely constrain system efficiency\cite{chen2019eyeriss},\cite{ yang2025efficient},\cite{ cheon20232941}.

Computing in-memory  (CIM) has emerged as a good solution to overcome these fundamental limitations \cite{ cheon20232941}, \cite{ guo202434}, \cite{jhang202422}. By performing computations directly within the memory array, CIM eliminates the energy-intensive data movement characteristic of von Neumann architectures. This approach enables massively parallel multiply-and-accumulate (MAC) operations through concurrent row activation, simultaneously achieving both high computational density and significant reduction in the number of data read/write operations between the processor and memory \cite{twins8T}. According to storage technology, CIM can be broadly categorized into two classes. One is Non-Volatile Memory-based CIM (NVM-CIM), such as Resistive Random-Access Memory (RRAM), Magnetoresistive Random-Access Memory (MRAM), Phase-Change Memory (PCM), etc. Although RRAM-based CIM can achieve high area efficiency due to its small bitcell size, it faces two main issues: the significant programmed error and stuck-at-fault ratio of RRAM devices considerably degrade network accuracy \cite{yang202533}. The primary challenges for MRAM-based CIM are its low on/off ratio and the difficulty of integrating MRAM arrays with complex CMOS peripheral circuits \cite{roy2020memory}. PCMs \cite{roy2020memory} offer high ON/OFF ratios, high density, and scalability, making them ideal candidates for crossbars. However, several technology challenges, such as conductance drift over time, high write energy and latency, and low endurance, can adversely affect the functionality of PCM-based CIM. The other class is based on volatile memory, such as Dynamic Random-Access Memory (DRAM) and static random-access memory (SRAM). The main drawback of DRAM-based CIM stems from DRAM itself, as it requires periodic refreshing of the stored weights because of facing the challenge of leakage and noise impact \cite{kim2023overview}, which significantly degrades the throughput of CIM.

Compared with the aforementioned memories, SRAM-based analog CIM (ACIM) architectures have demonstrated particular promise, delivering better improvements in energy efficiency and throughput while maintaining area efficiency\cite{zhang2017memory,zhang2023macc}, \cite{ chen202115}. Compared to Digital CIM (DCIM), ACIM demonstrates superior energy efficiency and reconfigurability, particularly in low-resolution scenarios \cite{ guo202428}. ACIM architectures can be broadly classified into three types \cite{kim2022overview}: voltage-domain, charge-domain, and current-domain. In voltage-domain CIM \cite{ KIMTCASIvoltagemode }, a shared read bitline (RBL) is driven by parallel pull-up or pull-down paths, which are activated based on the multiplication results of individual bitcells. However, this voltage-mode approach is often plagued by residual non-linearity and susceptibility to process variations. Charge-domain \cite{valavi201964} computing has been introduced to mitigate the non-linearity and variation issues inherent in analog computation. While this method enhances energy efficiency and throughput, it requires an additional switch per bitcell, an extra cycle for the accumulation operation, and suffers from a larger area footprint due to the necessary capacitors. In current-domain CIM \cite{kim65nm8TSRAMIMC},  it does not require a large number of capacitors in each bitcell, and the binary multiplication results from bitcells in the same column are accumulated by the voltage drop on the RBL. However, key challenges for current-domain ACIM persist in: (1) larger overhead analog-to-digital converter (ADC)\cite{kim2023neuro},\cite{yang2025high}, (2) excessive latency of multi-bit input\cite{kim20231},\cite{rajanna2021sram},\cite{saragada2024process}, (3) limited linearity of unit discharge current ($I_u$) of bitcell and dynamic range (DR) of voltage of read bitline ($V_{RBL}$) \cite{twins8T},\cite{kim65nm8TSRAMIMC}, (4) process-voltage-temperature (PVT) variation introducing non-negligible errors for MAC operations \cite{ kim65nm8TSRAMIMC}, \cite{ shaik2024impact},\cite{ KIMTCASIvoltagemode }.

This work proposes a multi-bit, highly reconfigurable CIM (1-7/2-4/1-7b, input/weight/output) by using bit-slicing (BS) with charge-sharing based binary weighted accumulator (BSCHA) and  reconfigurable (1-7b) in-memory analog to digital converter (IMADC) with shared references. The principal contributions of this work in addressing aforementioned challenges include:

(1) We propose a dual 8T bitcell architecture that enables ternary weights using a decoupled read path, integrated with a read word-line under-driven cascode (RWLUDC) technique. This technique achieves a 3.5× improvement in the DR of $V_{RBL}$ compared to prior work \cite{yu2024dual} and a 1.4× enhancement over conventional cascode structures, while simultaneously improving $I_u$ linearity by 7×.

\begin{figure}[t]
  \centering
  \includegraphics[width=1\linewidth]{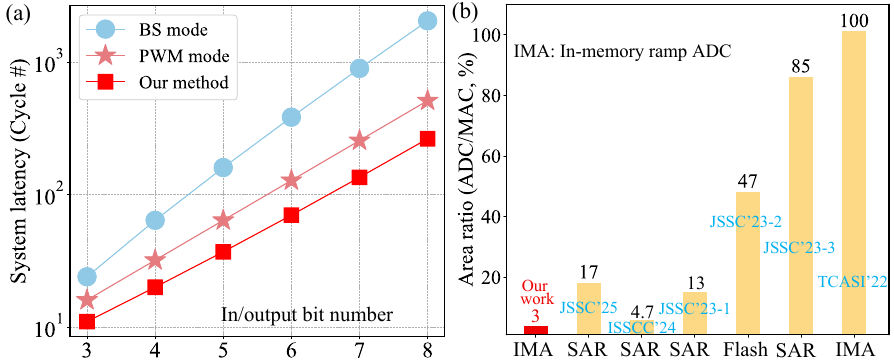}
  \caption{System latency and ADC area overhead:  (a) System latency (Our method, PWM mode \cite{dong2025topkima} and BS mode \cite{twins8T}. All three modes utilize the identical type of ADC, specifically the Ramp ADC.). (b) ADC area overhead (ADC area/MAC array area) of our work and previous works (JSSC'25 \cite{yin2025hybrid}, ISSCC'24 \cite{wang202434}, JSSC'23-1 \cite{yin2023cramming}, JSSC'23-2 \cite{lee202328}, JSSC'23-3 \cite{wang2023charge}, TCASI'22 \cite{KIMTCASIvoltagemode}). Note that SAR and Flash ADCs may have a faster conversion time depending on how many neurons share one ADC.}
  \label{latency_and_overhead}
\end{figure}

(2) Our work results in 1.9× and 6.6× improvement in latency over conventional pulse width modulation (PWM) mode and BS mode respectively under 7-bit input/output by using BSCHA as shown in Fig. \ref{latency_and_overhead}(a). It also increases MAC linearity by $\approx$23× compared to PWM mode.

(3) IMADC with shared references diminishes the hardware overhead of ADC without reducing throughput. Our ADC area overhead (ADC area/MAC array area) is merely 3\%, which is much lesser than prior work in Fig. \ref{latency_and_overhead}(b) (9× improvement compared to 27\% for traditional IMADC \cite{yu2024dual} and 1.5× improvement compared to 4.7\% for State-of-the-Art (SOTA) \cite{wang202434}). Post layout simulation results demonstrate the IMADC's robustness to both temperature fluctuations and process variations through replica biasing. The standard deviation of IMADC operation error increases by merely 1.21× at 70$^\circ$C and 1.18× in SS process corner compared to the TT corner at room temperature (27$^\circ$C).  

(4) We select three representative networks based on  MLP, VGG-8, and Vision Transformer (ViT) to evaluate the proposed method through quantization-aware training (QAT) and noise-resilient training (NRT). Under the influence of circuit non-idealities, these three models exhibit maximum accuracy degradations of 0.1\%, 0.4\%, and 0.4\%, respectively, compared to their ideal quantized models.

(5) The proposed macro demonstrates high energy/area efficiency  (1023.2 TOPS/W, 27 TOPS/mm\textsuperscript{2} at 1/2/1b) in 65 nm CMOS. It improves normalized energy efficiency by 1.4× and area efficiency by 2.6× over prior designs. Furthermore, system-level evaluation of VGG-8 on CIFAR-10, conducted using SPICE and NeuroSim, shows a normalized energy efficiency improvement of 6×.

The rest of this article is organized as follows. In Section \ref{sec:background}, we introduce the background and related works on ACIM SRAM macros and Section \ref{sec:Architecture and Methodology} describes the design details of hardware. Section \ref{Charge Sharing for Multi-bit} introduces analog weighted accumulator based on charge-sharing and software design. In Section \ref{sec:Result}, we show the results of hardware and software. Finally, we conclude our article in Section \ref{sec:Conclusion}.
\section{Background and Related Work}
\label{sec:background}
 
\subsection{Overhead of ADC for ACIM}
A critical bottleneck in analog CIM implementations lies in the tripartite trade-off between ADC-induced area overhead, energy consumption, and computational latency \cite{DACoverheadbig}.  For ACIM, the ADC is essential as it quantizes the MAC value, which is an analog signal. While column multiplexing techniques (sharing one ADC per $M$ columns) alleviate area costs \cite{DACSARADC,jiang2020c3sram}, they degrade throughput proportionally by $M\times$.  

To mitigate the overhead of conventional ADCs, an IMADC architecture with a conventional ramp generator has been proposed to directly convert analog dot-product results into digital codes \cite{kim65nm8TSRAMIMC},\cite{KIMTCASIvoltagemode},\cite{dong2025topkima}, \cite{dong2026cadc}. The initial ramp voltage is established by discharging one bitline ($V_{BL}$) after precharging, using $2^{n_o-1}$ replica bitcells for an $n_o$-bit resolution ADC, which are also employed in calibration. Subsequently, $2^{n_o}$ replica bitcells dedicated to ramp generation modulate the complementary bitline voltage linearly to generate reference voltage levels. A sense amplifier compares the differential reference voltage  ($V_{BL}$-$V_{BLB}$). In each cycle, the sense amplifier produces a thermometer code based on the comparison between the dot-product output and the current reference level \cite{dong2025topkima}. This has similar or better throughput as earlier designs\cite{DACSARADC,jiang2020c3sram} but with the advantage of reconfigurability (of resolution and MAC size) as well as robustness to PVT variations due to replica biasing. However, such implementations are limited to binary neural networks (BNNs), which exhibit significant accuracy degradation when executing complex neural architectures \cite{kim65nm8TSRAMIMC}, \cite{KIMTCASIvoltagemode}. Also, The per-column ADC integration imposes substantial hardware overhead that scales exponentially with resolution $n_o$, requiring $2^{n_o}$ additional bitcells per column (e.g., 64 cells for $n_o=5$). Fig. \ref{latency_and_overhead}(b) shows ADC area overhead defined as the ratio of ADC area over MAC array area in previous works\cite{yin2025hybrid,yu2024dual,zhang2023macc,lee202328,wang2023charge,KIMTCASIvoltagemode}.  

\subsection{Design Challenges of multi-bit input}
BS is a widely-adopted technique for enhancing input bit resolution ($n_i$) through sequential bit application and post-digitization partial sum (PSUM) accumulation \cite{twins8T}, \cite{saragada2024process}.  However, when employing a ramp ADC for output quantization, the total latency of BS required becomes $n_i\times 2^{n_o}$ clock cycles, since each bit demands $2^{n_o}$ clocks. Moreover, this incurs $n_i$ times energy overhead of ADC as well as additional digital circuits for combining PSUMs.  The principle of the PWM input scheme is that the input value is proportional to the pulse width. Given an input resolution of ${n_i}$ bits, the maximum representable value is $2^{n_i}$, requiring up to $2^{n_i}$ clock cycles to encode. Therefore, for multi-bit input, PWM-based designs \cite{dong202015,gonugondla201842pj,jhang2021challenges} incur a total time cost of ($2^{n_i}$ + $2^{n_o}$) clock cycles, where $2^{n_i}$ accounts for the input pulse duration and $2^{n_o}$ for the output ADC conversion. Fig. \ref{latency_and_overhead}(a) shows this tradeoff between system latency and input/output bit resolution (assuming $n_i=n_o$).

Another issue with the PWM approach for multi-bit input is the signal margin, which can be analyzed from the perspective of DR per MAC operation (DR/MAC)  \cite{kim65nm8TSRAMIMC}. The MAC operation $M A C=\sum_{k=1}^N w_k x_k$. Compared to the bit-slicing method ($x$ is a single bit), the input $x$ (multi-bit) in the PWM scheme is inherently larger in magnitude. This leads to a significantly larger MAC result, which consequently substantially reduces the signal margin.

\subsection{Limited DR of \texorpdfstring{$V_{RBL}$}{VRBL}}
While employing multiple read wordlines (RWLs) operations improves energy efficiency and throughput, this technique inherently expands the MAC output voltage range. Consequently, under the fixed DR  of $V_{RBL}$, the signal margin (the minimum voltage difference between successive MAC operations \cite{jhang2021challenges}) deteriorates significantly \cite{kim65nm8TSRAMIMC}. A smaller signal margin leads to higher quantization errors in the ADC. Therefore, increasing the DR of $V_{RBL}$ is imperative to mitigate this issue.  A source degeneration architecture \cite{twins8T} that simultaneously enhances the signal margin and extends the usable voltage swing range is proposed. This solution demonstrates a 1.44× improvement in signal margin compared to conventional 6T-SRAM cells while preserving MAC operation precision.

\subsection{Quantization-aware training}
To preserve system-level inference accuracy when deploying networks on the proposed hardware in the presence of practical non-idealities (hardware noise and transistor mismatch), we adopt quantization-aware training (QAT) as an algorithm–hardware co-optimization step. Quantization-aware training is widely recognized as a key enabler for deploying deep neural networks on resource-constrained hardware platforms \cite{li2016ternary}. By explicitly modeling the quantization of weights and activations during training, QAT substantially alleviates the accuracy degradation (10\%–80\% in \cite{han2025mitigating}) that can arise with post-training quantization (PTQ). This joint optimization allows the model to adapt to low-precision representations, reducing memory footprint and simplifying arithmetic operations \cite{yin2020xnor,dong2022backpropagation}. As a result, the trained quantized networks can be mapped directly onto integer-based accelerators, delivering improvements in throughput and energy efficiency. Overall, QAT helps bridge algorithm design and practical hardware deployment, ensuring that hardware-level performance gains translate into robust system-level accuracy.

\section{Proposed Hardware Architecture and Methodology}
\label{sec:Architecture and Methodology}
\subsection{Hardware Overall}\label{Hardware Architecture}

\begin{figure}[t]
  \centering
  \includegraphics[width=0.85\linewidth]{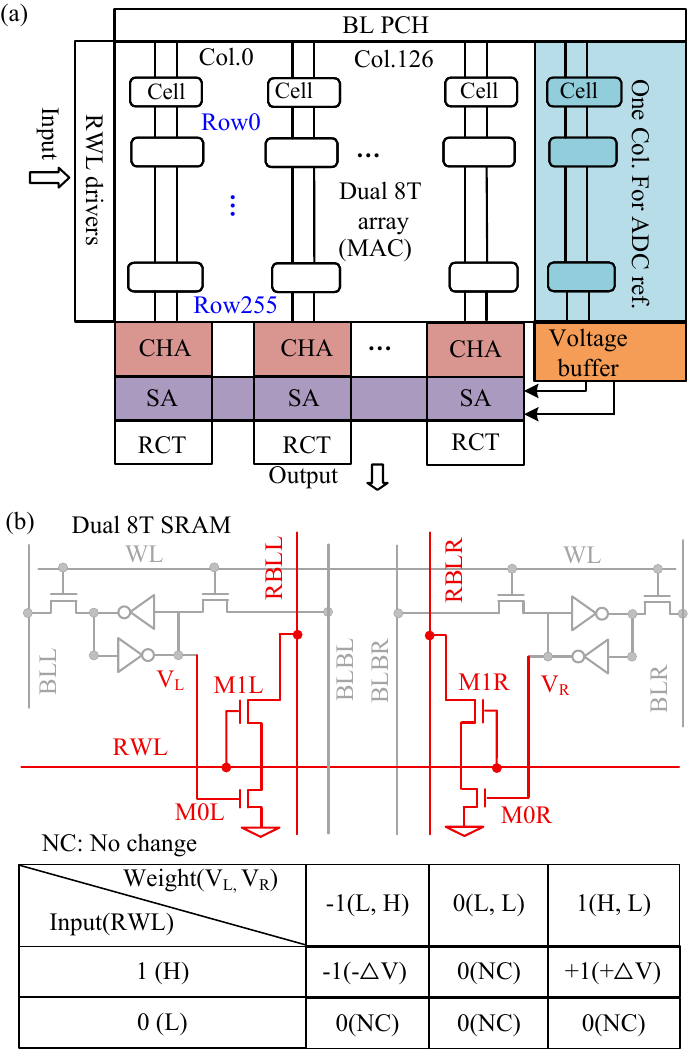}
  \caption{(a) Hardware structure of the proposed dual 8T SRAM CIM and IMADC. (b) Information of dual 8T SRAM bitcell.}
  \label{Overall_Hardware_Architecture}
\end{figure}

The top-level schematic of the SRAM-based CIM and IMADC is illustrated in Fig. \ref{Overall_Hardware_Architecture}. The schematic includes a $256\times 127$ array of dual 8T SRAM bitcells that perform MAC operations, along with a single reference column ($256\times 1$) that provides reference voltages for the reconfigurable ADC and is shared across the 127-column MAC array. The design also incorporates charge-sharing-based binary-weighted accumulators (CHAs), sense amplifiers (SAs), ripple counters (RCTs), input drivers, and voltage buffers.

\begin{figure*}[t]
  \centering
  \includegraphics[width=0.9\linewidth]{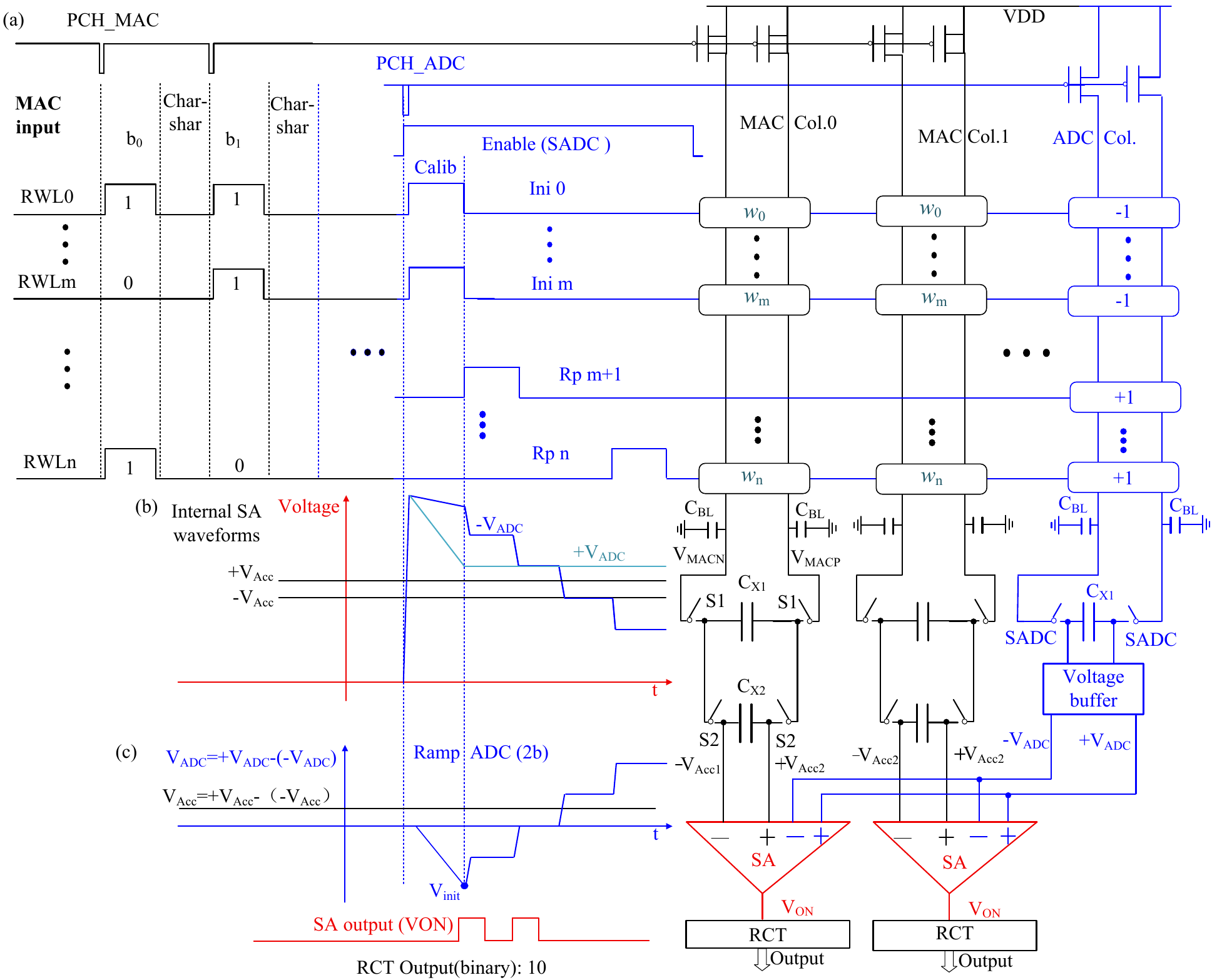}
  \caption{(a) Detailed circuit and timing diagram of MAC columns and IMADC. (b) Internal signals of SA. (c) IMADC principle.}
  \label{timing_diagram}
\end{figure*}

To clearly illustrate the system operation, Fig. \ref{timing_diagram} (a) depicts the timing sequences for both MAC operations and ADC conversion. During operation, the input bits $b_i$ are applied serially to generate the differential voltage $V_{MAC}=V_{MACP}-V_{MACN}=\sum_{k=1}^N W_k X_k^i$ on the bitlines through current-mode operation, where $N=256$ represents the dimension of the input vector. The near-memory capacitors $C_{X1}$ and $C_{X2}$, which have equal capacitance values, perform binary-weighted accumulation of partial sums in charge mode to produce the final accumulated voltage $V_{Acc}(+V_{Acc}-(-V_{Acc}))$. Then, the ADC column generates a reference ramp voltage $V_{ADC} (+V_{ADC}-(-V_{ADC}))$ that increments every clock cycle. This ramp voltage is shared with all remaining columns, enabling each of the 127 SAs to simultaneously compare it with their corresponding $V_{Acc}$ values. Because the ramp voltage is shared across multiple columns, differential voltage buffers are implemented to enhance its driving capability. The RCTs then convert the thermometer-coded outputs from the SAs into binary format. Fig. \ref{timing_diagram}(b) shows the signal waveforms at the four input terminals of the SA, while Fig. \ref{timing_diagram}(c) illustrates the quantization process of the 2-bit ramp ADC and the output of the SA.

\begin{figure}[t]
  \centering
  \includegraphics[width=1.0\linewidth]{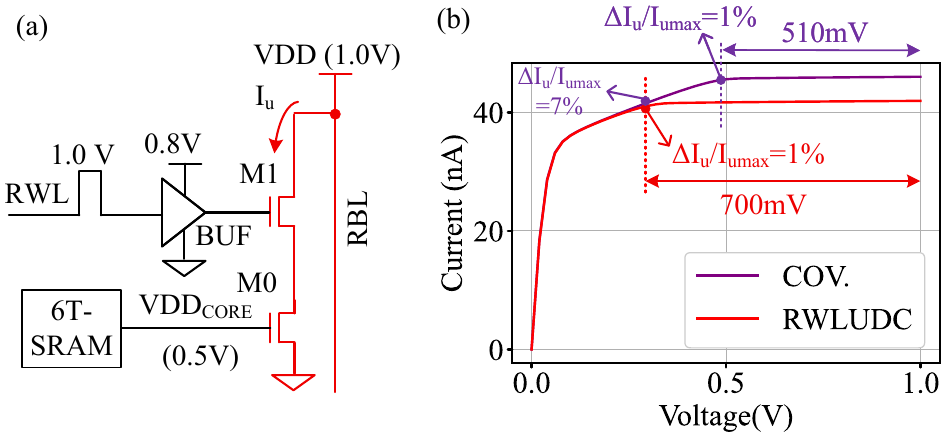}
  \caption{ (a) Schematic of RWLUDC. (b) Voltage swings and current variations between the proposed RWLUDC and conventional cascode architecture based on SPICE  simulation.}
  \label{RWLUDC}
\end{figure}
\subsection{In-Memory Computing Using Dual 8T SRAM Bitcell}
\label{CIM MAC}

The proposed macro architecture employs a dual 8T SRAM bitcell, as illustrated in Fig. \ref{Overall_Hardware_Architecture}(b). This design enables binary multiplication between input data and ternary weights through an decoupled read pathway implemented by four red NMOS transistors (arranged symmetrically as depicted in the schematic).

The multiplication mechanism operates as follows: The read wordline (RWL) receives binary input signals represented by either high or low DC voltage levels. Meanwhile, each 6T-SRAM bitcell stores ternary weight values encoded in three possible states: -1 (V$_L$=low, V$_R$=high), 0 (both V$_L$ and V$_R$=low), or 1 (V$_L$=high, V$_R$=low). Prior to computation, both read bitlines (RBLL and RBLR) are initialized to high potential through PMOS pre-charge transistors controlled by the \textit{PCH\textunderscore MAC} signal (Fig. \ref{timing_diagram}).

The multiplication truth table appears in Fig. \ref{Overall_Hardware_Architecture}. Key observations include: (1) With RWL at logic low (input=0), no current discharges the
RBLL or RBLR, regardless of the weight stored in the 6T-SRAM bitcell.  (2) When RWL activates (input=1), selective discharge occurs through the NMOS transistor, where the gate is linked to the internal 6T-SRAM cell with a high voltage (0.5 V), creating either positive or negative differential voltage between RBLL and RBLR (+$\Delta$V/-$\Delta$V). This voltage difference encodes the multiplication result. When the weight is 0 both V$_L$ and V$_R$=low), no current discharges the RBLL or RBLR, regardless of the inputs, which is called zero-skipping (ZOSKP) for weights. Accumulation is achieved through the simultaneous discharge of the bitline capacitor $C_{BL}$ by all bitcells within a column. The resulting MAC value exhibits a linear dependence on the voltage drop generated during this capacitive discharge process. A comprehensive analysis of this mechanism will be presented in Section \ref{Charge Sharing for Multi-bit}.


\subsection{RWL Underdrive enabled Cascode for enhanced DR of \texorpdfstring{$V_{RBL}$}{VRBL}}
\label{Dual 8T RWLUDC-based SRAM bitcell}
A fundamental limitation in current-based MAC architectures stems from the nonlinear discharge characteristics of the bitline capacitance $C_{\text{BL}}$. This nonlinearity arises from the dependence of the unit bitcell discharge current $I_u$ on the drain-source voltage $V_{\text{DS}}$, which varies with the dynamically changing bitline voltage \cite{KIMTCASIvoltagemode}. This can be improved by increasing the output impedance of the effective current source created by transistors M0 (L/R) and M1 (L/R) in Fig. \ref{Overall_Hardware_Architecture} by making M1 behave as a cascode\cite{AllenHolberg}. We propose to achieve this by RWL underdrive voltage enabled cascode (RWLUDC) technique--the supply voltages for RWL is set to $0.8$ V while RBLL/RBLR pre-charging are set at $1$ V while gate voltage of M0 is at $0.5$ V (6T-SRAM supply voltage). The minimum RBL voltage for which M1 acts as a cascode is given by $V_{min}=V_{RWL}-V_{T1}$ where $V_{T1}$ denotes the threshold voltage of M1, i.e. lower $V_{RWL}$ leads to higher RBL swing. After lowering $V_{RWL}$, the saturation region range of the Cascode (from $V_{min}$ to $V_{DD}$) increases because $V_{min}$ decreases. From the simulation results in Fig.~\ref{RWLUDC}(b), the saturation region of the RWLUDC is 190 mV larger than that of the conventional cascode structure. We implemented the RWLUDC by reducing the power supply voltage of the RWL buffer to 0.8 V instead of $V_{DD}=1.0$ V as illustrated in Fig.~\ref{RWLUDC}(a). The cost associated with this approach is the requirement for an additional 0.8 V DC power supply to provide power to the buffer.

The proposed technique achieves a significant improvement in voltage swing characteristics, as demonstrated in Fig.~\ref{RWLUDC}. For a 1\% variation in $I_u$, we obtain an RBL DR of 700 mV, which represents a 7$\times$ reduction in current variation sensitivity compared to conventional implementations. In contrast, conventional approaches must constrain their DR to approximately 510 mV (1.4$\times$ reduction) to maintain the same 1\% $I_u$ variation tolerance. This performance limitation becomes even more pronounced in single-transistor (7T bitcell) discharge paths, where the achievable signal swing is further restricted to approximately 200 mV \cite{yu2024dual}. 

We simulated the timing of RWL at 0.8V and 1V, and found that the latency of RWL at 0.8V increases by 0.25 ns. Our maximum frequency is 200 MHz, corresponding to a cycle time of 5 ns. The additional delay of 0.25 ns accounts for only 5\% of the cycle (0.25 ns/5 ns = 5\%), which is negligible and will not affect the overall system throughput.

\subsection{Reconfigurable 1-7 bit In-Memory Analog-Digital Converter (IMADC) with Shared References}\label{shared reconfigurable IMADC}
Fig. \ref{timing_diagram} illustrates the architecture and operational sequence of the IMADC, highlighted in blue. The IMADC employs replica cells that are structurally identical to those used in the MAC array. The conversion process consists of two consecutive phases: (1) the initial ramp voltage $V_{init}$ is established within a single clock cycle by activating $2^{n_o-1}$ bit-cells assigned a weight of $-1$ as shown in Fig. \ref{timing_diagram}; (2) subsequently, the ramp is generated by sequentially enabling $2^{n_o}$ bit-cells with a weight of $+1$ at each clock cycle, followed by successive comparisons using a SA. The schematic in Fig. \ref{timing_diagram} outlines the global ramp reference voltage ($V_{ADC}$) generation for a 2-bit IMADC and the quantization process of the accumulated voltage ($V_{Acc}$) in the MAC column.

Our IMADC design fundamentally differs from prior architectures through a shared reference scheme. Prior implementations such as \cite{KIMTCASIvoltagemode} employed per-column ramp generators, incurring an area overhead of approximately 50\%. In our design, this is replaced by a single global reference generator. To prevent overlap between MAC and ADC operations and minimize power dissipation from switching and voltage buffers, an \textit{Enable} signal (\textit{SADC}) is introduced to ensure the ADC starts after the MAC operation ends. \textit{SADC} is used to gate the \textit{PCH} and \textit{RWL} to create \textit{PCH\textunderscore ADC} and \textit{RP} signals for the ADC column and it also controls the activation of the SA. The number of cells engaged in ramp generation can be configured according to the target resolution. With $N=256$ rows available, a maximum resolution of $n_o=7$ bits is achievable, utilizing 128 cells; the remaining 128 cells are reserved for the initial ramp voltage and calibration purposes to ensure the ramp crosses zero, as discussed in \cite{KIMTCASIvoltagemode}.

While the proposed approach achieves a significant reduction in area overhead to $\approx 3\%$ (Fig. \ref{latency_and_overhead}(b)), it requires two modifications--(1) the SA has to be converted to a double differential one (Schematic of SA is shown in Fig. \ref{timing_diagram}) \cite{marisa2017pseudo} that compares the differential ramp voltage $V_{ADC}$ with a differential $V_{Acc}$ and (2) a voltage buffer has to be used after the ramp generator to drive the input capacitance of $127$ SAs. The output signal $V_{ON}$ of the SA is the input of the RCT. In the scenario where $V_{Acc}=2$, as illustrated in Fig. \ref{timing_diagram} (c), the output of the RCT is $10$. The RCT acts as a thermometer to binary code converter, eliminating the need for large register arrays to store the thermometer code, thereby improving  overall area efficiency.

\begin{figure}[t]
  \centering
  \includegraphics[width=0.95\linewidth]{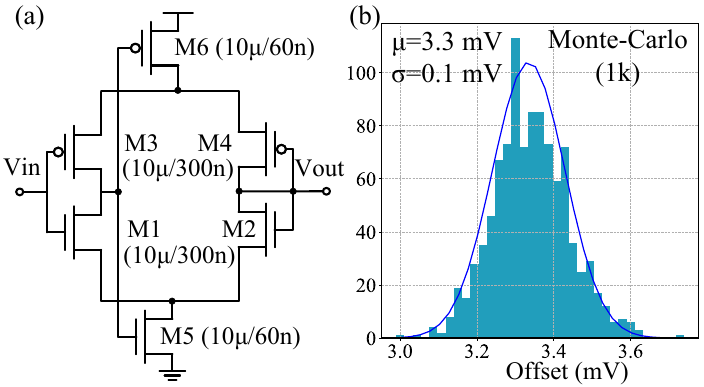}
  \caption{Self-biased voltage buffer (a) Circuit Diagram. (b) Monte
Carlo simulation for voltage offset showing it to be less than ADC LSB.}
  \label{voltage buffer and ADC timing}
\end{figure}

The IMADC incorporates two self-biased voltage buffers (Fig. \ref{voltage buffer and ADC timing}(a)).  The principle of the self-biased voltage buffer is as follows. When the input voltage $V_{in}$ increases, the drain voltages of $M_1$ and $M_3$ decrease. This triggers $M_6$ to turn on, causing a surge in the branch current. This current flows through $M_4$ to the output node, effectively charging the capacitive load (comprising 127 SAs). Under this condition, the current through $M_5$ remains near zero. Conversely, when $V_{in}$ decreases, $M_5$ becomes highly conductive, allowing a large current to be sunk from the output capacitor through $M_2$ to ground. This configuration allows the buffer to provide high sourcing and sinking peak currents on demand during transitions, while maintaining a very low quiescent (static) current during steady-state.

The proposed self-biased buffer achieves a 62 MHz bandwidth and an 88V/µs slew rate with a dynamic power consumption of only 40 µW from a power supply of $1$ V. With an ADC resolution of 4.8 mV, the settling time is merely 0.05 ns even when driving the input capacitance ($\approx$ 100 fF ) of 127 SAs simultaneously. Consequently, even if the ADC operates at its maximum frequency of 200 MHz (with a 5 ns clock period), the settling time accounts for only 1\% of the cycle, indicating a negligible impact on performance.

In the Monte Carlo (MC) simulation depicted in Fig. \ref{voltage buffer and ADC timing}(b), the offset voltage determined by the output difference of the two voltage buffers minus the input difference is shown. 
The MC simulation results (Fig. \ref{voltage buffer and ADC timing}(b)) show the offset voltage distribution, defined as the output voltage difference minus the input voltage difference of the two buffers. The results indicate a mean offset of 3.3 mV with a standard deviation of 0.1 mV, which is well below the ADC's LSB value of 4.8 mV, thereby eliminating the need for offset calibration. Despite the inclusion of these buffers and a more complex SA, the proposed architecture consumes approximately 3\% less energy compared to the ramp generator in every column\cite{KIMTCASIvoltagemode} owing to reduced bitline capacitances and lower switching energy achieved through the shared reference scheme.

Comparing traditional analog buffers, such as a super source follower (SSF), our benefits are as follows. 1) Minimized input-output voltage drop: A major drawback of the SSF is the inherent $V_{GS}$ drop between the input and output, which typically ranges from 200 mV to 500 mV in 65 nm CMOS technology. Such a large drop significantly compresses the signal dynamic range, which is unacceptable for high-precision reference sharing. In contrast, our proposed buffer achieves a minimal input-output voltage offset of only 3.3 mV, ensuring nearly unity gain and maximizing the available voltage for the SAs. 2) Standard buffers often require complex external biasing networks to maintain stability and bandwidth. The proposed architecture utilizes a self-biasing mechanism, which not only reduces the overall silicon area and routing complexity but also allows the buffer to automatically adapt its power consumption based on the input level.

We performed parasitic extraction for layout, and the equivalent total parasitic capacitance of the ADC reference signal is 96 fF, with a parasitic resistance of 160 ohms. After incorporating the parasitic parameters into the circuit for post-layout simulation,  the settling time skew between the nearest column 1 and the farthest column 127 is approximately 0.1 ns, which is negligible compared to our fastest clock frequency of 200 MHz.

\subsection{Method for multi-bit weight}
\label{Method for multi-bit weights}
\begin{figure}[t]
  \centering
  \includegraphics[width=1\linewidth]{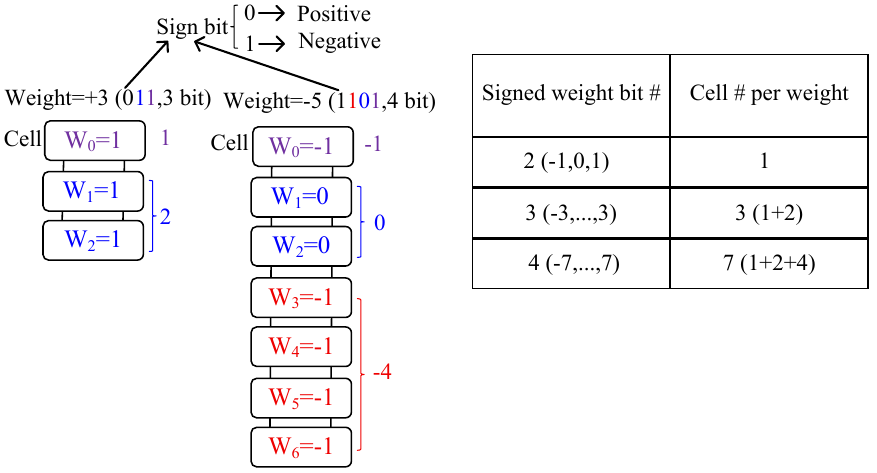}
  \caption{Method for multi-bit weight with two cases (3-bit and 4-bit weights).}
  \label{multi-bit_weight}
\end{figure}

Our hardware architecture supports multi-bit weights to cater to the requirements of complex networks. Multiple-bit weights can be realized by connecting multiple cells in parallel.  Fig. \ref{multi-bit_weight} illustrates how multiple cells are used to represent a multi-bit weight in both cases (3-bit and 4-bit weights). For instance, in a 4-bit configuration, excluding the sign bit, the remaining three bits can be implemented using 1, 2, or 4 cells in parallel (for a total of 7 cells for one 4-bit weight). The sign of the weight can be intrinsically implemented using a dual 8T architecture based on the usage of the left or right cells. The inputs of 7 cells are identical.  Although this approach  require a significant number of cells, especially as the bit count increases, the advantages of this method include low latency—requiring only one cycle for implementing multi-bit weight—along with reconfigurability and high flexibility, as it can support weights ranging from 2 to 4 bits.  It is worth noting that this method is primarily suitable for neural networks with low-precision weights. As the weight precision increases, the number of cells required to represent a single weight grows exponentially, which inevitably leads to an increased system level overhead.


\section{Analog Weighted Accumulator and software design}
\label{Charge Sharing for Multi-bit}
\subsection{Operating Principle of Analog Accumulator}
Conventional BS in CIM applies the input bits $b_0$ to $b_{n_i-1}$ serially over $n_i$ cycles while the MAC voltage is digitized separately for every bit to produce psums $P_0$ to $P_{n_i-1}$. The final MAC result is obtained by weighted summation in the digital domain as $P_{n_i-1}+\frac{P_{n_i-2}}{2}+...+\frac{P_0}{2^{n_i-1}}$.
The resulting energy ($E_{conv}$) and latency ($T_{conv}$) in terms of clock cycles of the conventional BS approach are given by:

\begin{equation}
\label{eq:bs_conv}
\left\{\begin{array}{l}
E_{c o n v}=n_i\left(E_{M A C}+E_{A D C}+E_{d i g}\right) \\
T_{c o n v}=n_i\left(T_{M A C}+T_{A D C}+T_{d i g}\right)
\end{array}\right.
\end{equation}

where $E(T)_{MAC}$, $E(T)_{ADC}$ and $E(T)_{dig}$ denote the energy (latency) of MAC, ADC and digital weighted accumulation respectively for every cycle. This incurs a heavy penalty in terms of energy and latency due to the repeated operation of the ADC over $n_i$ cycles.

To overcome this issue, we propose a binary weighted accumulation before digitization technique accomplished by the addition of two capacitors and four switches between RBLL and RBLR (Fig. \ref{timing_diagram}). The simplified circuit diagram is illustrated in Fig. \ref{charge sharing}(a), where $C_{BL}$ represents the parasitic capacitance on RBL, and $I_{MACN}$ and $I_{MACP}$ are the total discharge currents of each column. By applying Kirchhoff's Current Law, we can get Eq. \eqref {eq:IMACPN}. 
\begin{equation}
\left\{\begin{array}{l}
I_{\mathrm{MACN}}+I_{C B L N}+I_{C X N}=0 \\
I_{\mathrm{MACP}}+I_{C B L P}+I_{C X P}=0
\end{array}\right.
\label{eq:IMACPN}
\end{equation}
From the capacitor's current and voltage formulas, we can obtain Eq. \eqref {eq:IMACP/N} from Eq. \eqref {eq:IMACPN}.
\begin{equation}
\left\{\begin{aligned}
I_{M A C N} & =-I_{C X N}-I_{C B L N} \\
& =C_{X 1} \frac{d\left(V_{M A C P}-V_{M A C N}\right)}{d t}-C_{B L} \frac{d V_{M A C N}}{d t} \\
I_{M A C P} & =-I_{C X P}-I_{C B L P} \\
& =-C_{X 1} \frac{d\left(V_{M A C P}-V_{M A C N}\right)}{d t}-C_{B L} \frac{d V_{M A C P}}{d t}
\end{aligned}\right.
\label{eq:IMACP/N}
\end{equation}
Subtracting the two currents in Eq. \eqref {eq:IMACP/N} yields Eq. \eqref {eq:IMACP-N}.
\begin{equation}
\begin{aligned}
& I_{\mathrm{MACP}}-I_{\mathrm{MACN}}=\left(2 C_{X 1}+C_{B L}\right) \frac{d\left(V_{M A C N}-V_{M A C P}\right)}{d t} \\
\end{aligned}
\label{eq:IMACP-N}
\end{equation}

Intuitively, $C_{X1}$ is used to sample the MAC voltage in every cycle for every input bit, and then redistribute it with an \textbf{equally sized} capacitor $C_{X2}$. The bitcell array layout is designed to utilize metal layers M1 to M4. To optimize the overall chip area, metal layers M6 to M7 are allocated for constructing MOM capacitors $C_{X1}$ and $C_{X2}$, while metal layer M5 is grounded to shield the bitcell array from the capacitors. Each capacitor has an area of $3\times 47 \mu m^2$ , enabling the straightforward placement of two MOM capacitors on top of every column in the bitcell array without incurring any additional area overhead.

This process results in the binary weighted charge sharing based accumulation required to combine the partial sums for every bit. It is similar in principle to charge redistribution DACs\cite{AllenHolberg}; however, instead of an input digital code, a different analog MAC voltage is input in every cycle.

The detailed process is shown in Fig. \ref{charge sharing}(b) including the timing diagram of pre-charging, input pulse, and charge-sharing control switches S1 and S2. We analyze the case for the \textit{i}-th bit of the input (assuming both capacitors were reset before the first input bit was presented):\\
1) During pre-charging, switch S1 is closed, and S2 is open. After pre-charging, the voltages of the two $C_{BL}$s and the voltages across $C_{X1}$ reach VDD resetting the previously stored voltages.\\
2) Subsequently, RWL is enabled for time $\Delta t$, initiating the discharge of $C_{BL}$. The voltage on $C_{X1}$ changes as the two $C_{BL}$s discharge.  We can derive Eq. \eqref {eq:Vmac} from Eq. \eqref {eq:IMACP-N} for the voltage $V_{MAC}=V_{MACN}-V_{MACP}$ as:
\begin{equation}
\begin{aligned}
& V_{MAC}^i=\frac{\left(I^i_{\text {MACP}} - I^i_{\text {MACN }}\right) \Delta t}{2 C_{X1}+C_{BL}}=\frac{I_u \sum_{k=1}^N W_k X^i_k}{2 C_{X1}+C_{B L}} \Delta t
\end{aligned}
\label{eq:Vmac}
\end{equation}
where $V_{MAC}^i$ denotes the MAC voltage developed on the capacitors $C_{BL}$ and $C_{X1}$ when the i-th input bit is presented to the circuit. A simplified equivalent circuit is presented in Fig. \ref{charge sharing}(b) (Step 1).

3) Subsequently, S1 opens, and S2 closes, initiating the charge-sharing  between capacitors $C_{X1}$ and $C_{X2}$ as shown in Fig. \ref{charge sharing}(b) (Step 2). The voltage on capacitor $C_{X2}$ denoting the final accumulated voltage $V_{Acc}$ becomes a combination of half of its previous value $V_{Acc}^{i-1}$ and half of the the new MAC value $V_{MAC}^{i}$ as shown in in Eq. \eqref {eq:Vchsh}: 

\begin{figure}[t]
  \centering
  \includegraphics[width=0.9\linewidth]{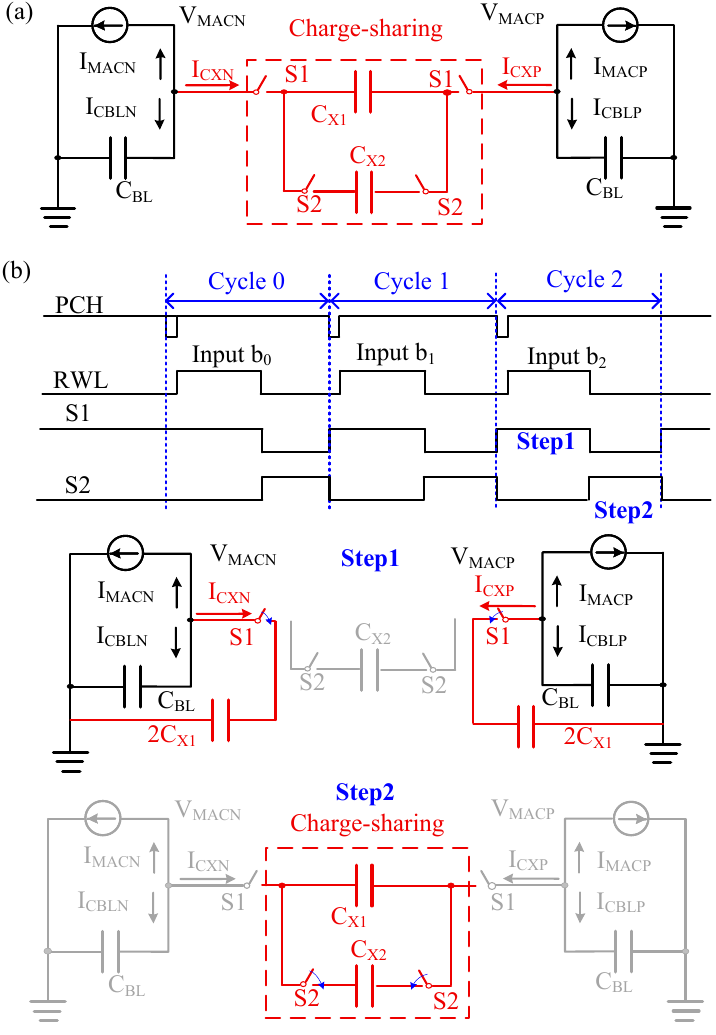}
  \caption{(a) Effective circuit model for generation of differential MAC voltage and charge sharing. (b) Timing diagram of charge sharing and equivalent circuit for two steps. The MAC voltage is sampled on $C_{X1}$ when S1 and RWL are enabled (Step 1), while charge redistribution between $C_{X1}$ and $C_{X2}$ occurs when S2 is enabled (Step 2).}
  \label{charge sharing}
\end{figure} 
\begin{equation}
\left\{\begin{aligned}
& C_{X2}V_{Acc}^{i-1}+C_{X1}V_{MAC}^{i}=V_{Acc}^i\left(C_{X1}+C_{X2}\right) \\
& V_{Acc}^i=\frac{V_{Acc}^{i-1}}{2}+\frac{V_{MAC}^{i}}{2}=\sum_{k=0}^{i}\frac{V_{MAC}^k}{2^{i+1-k}}
\end{aligned}\right.
\label{eq:Vchsh}
\end{equation}

The energy and latency for our approach can be written as:

\begin{equation}
\label{eq:bs_prop}
\left\{\begin{aligned}
E_{\text {proposed }} & =n_i\left(E_{M A C}+E_{\text {ana }}\right)+E_{A D C} \\
T_{\text {proposed }} & =n_i\left(T_{M A C}+T_{\text {ana }}\right)+T_{A D C}
\end{aligned}\right.
\end{equation}

where $E(T)_{ana}$ denotes the energy (latency) of analog weighted accumulation and other terms are as defined earlier. The major advantage of our approach stems from the fact that the ADC is now operated only once as opposed to $n_i$ times as in Eq. \eqref{eq:bs_conv}. The errors induced by the analog accumulation can be compensated by co-design of the algorithm as shown in Sec. \ref{noise training}.

In contrast to the PWM mode where the input latency is $2^{n_i}$, our method demonstrates clear advantages in terms of input latency, as shown in Fig. \ref{latency_and_overhead}(b) (18× improvement for 7-bit input). Furthermore, in comparison to the PWM mode for multi-bit input, our method exhibits superior linearity in generation of $V_{Acc}$ due to smaller voltage swing on  $V_{RBLL/R}$. As mentioned in Section \ref{Dual 8T RWLUDC-based SRAM bitcell}, Early effect of discharge transistor of the bitcell (M0, M1) results in a decrease in $I_u$, subsequently diminishing $V_{Acc}$ and giving rise to considerable errors.

\subsection{Non-ideality Analysis}

\begin{figure}[t]
  \centering
  \includegraphics[width=\linewidth]{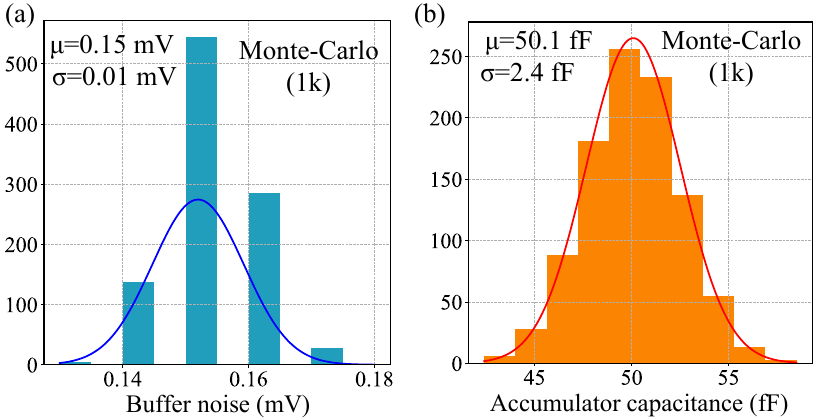}
  \caption{ (a) Monte Carlo simulation for buffer noise. (b) Monte Carlo simulation for capacitance of analog accumulator.}
  \label{noise_mismatch}
\end{figure}

\textbf{(1) Noise analysis.} Compared to other charge sharing methods\cite{DACSARADC}, our proposed method does not require binary weighted capacitors, reducing area and matching requirements. Usage of differential architecture helps to reduce the effect of charge injection in the switches. In our implementation, $C_{X1}=C_{X2}=50$ fF is chosen to make it approximately similar to $C_{BL}/2$ (balancing two parts in denominator of Eq. \eqref{eq:Vmac}) since it provides a good tradeoff between reducing thermal noise and maintaining large enough MAC voltages.
Thermal noise comprises two primary components: sampling noise originating from the accumulator and the shared reference, and buffer noise associated with the shared reference. The contribution of sampling noise from each switch is kT/C (k is the Boltzmann constant and T is the absolute temperature in Kelvin), which is 20 $\mu$V in our design given $C_{X1}=C_{X2}=50$ fF. As the accumulator operates over $n_i$ cycles, the total sampled noise is $\sum_{k=0}^{n_i-1} \frac{V_{noise(k)}}{2^{n_i-k}}$, which is dominated by the contribution from the final cycle, as noise from earlier cycles is attenuated by a factor of 1/2 during the accumulation process. For a 5-bit input system, the IMADC LSB (step size of IMADC) is 4.8 mV, while the total noise from the four switches, calculated by power summation due to their uncorrelated nature, is 40 $\mu$V. This indicates that the thermal noise is negligible.

We performed a noise simulation on a buffer with a load capacitance of 100 fF. The noise was integrated over the frequency range of 1-1000 MHz to obtain the effective noise voltage value. MC simulations were performed on this effective noise voltage value, yielding the results presented in Fig. \ref{noise_mismatch}(a), which show a mean noise value of 0.15 mV. This value is significantly smaller compared to the 4.8 mV (LSB of the IMADC), demonstrating negligible buffer noise impact on the quantization precision of IMADC.

\textbf{(2) Mismatch analysis of analog accumulator.} 
\begin{figure}[t]
  \centering
  \includegraphics[width=\linewidth]{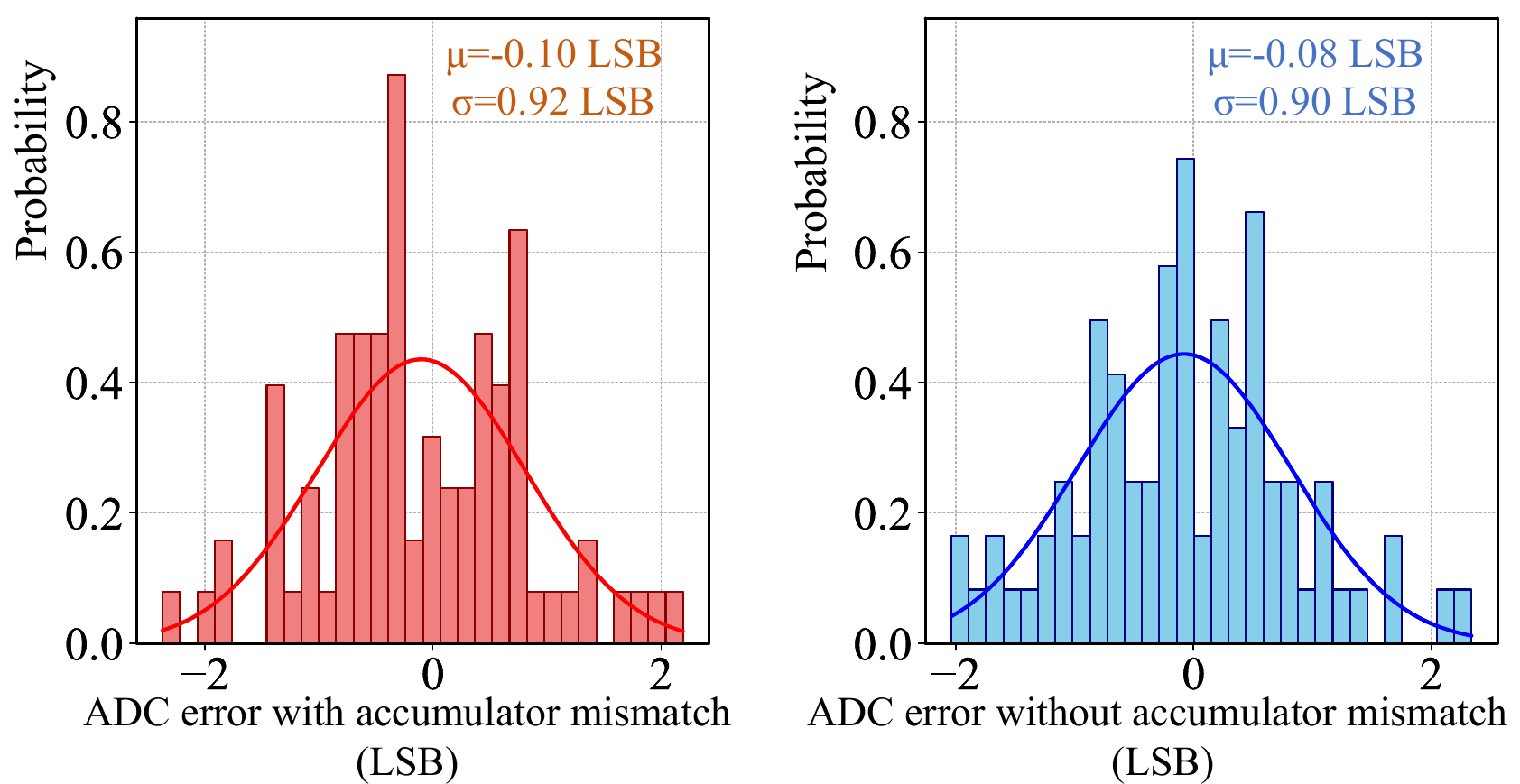}
  \caption{ADC error distribution (ADC error= (simulated ADC output-theoretical ADC output)/resolution.}
  \label{cap_mismatch_adc_error}
\end{figure}
Considering the impact of capacitor mismatch on charge sharing, we conducted MC simulations on the accumulator capacitance values, with results as shown in the Fig. \ref{noise_mismatch}(b). The capacitance values follow a normal distribution \textit{N}(50.1 fF, 2.4 fF). Based on the $3\sigma$ rule, $C_{X2} = \mu + 3\sigma$ = 57.3 fF  are used for SPICE simulation while $C_{X1} $=50 fF. We simulated the actual CIFAR-10 dataset on VGG-8 model under a 4-bit ADC and 2-bit weight configuration, and the resulting ADC error distribution is shown in Fig. \ref{cap_mismatch_adc_error}(left). Based on the distribution of MAC values in the quantized network, the step size of the ADC is 16. For comparison with the effect of mismatch, we also plotted the ADC error distribution without accumulator mismatch in Fig. \ref{cap_mismatch_adc_error}(right). The standard deviation varied by only 2\%. To further validate the impact of capacitor mismatch on network accuracy, we incorporated a noise model with mismatch into the VGG-8 network. The inference accuracy decreased by merely 0.5\% compared to the same network utilizing  noise model without mismatch.

\begin{figure}[t]
  \centering
  \includegraphics[width=\linewidth]{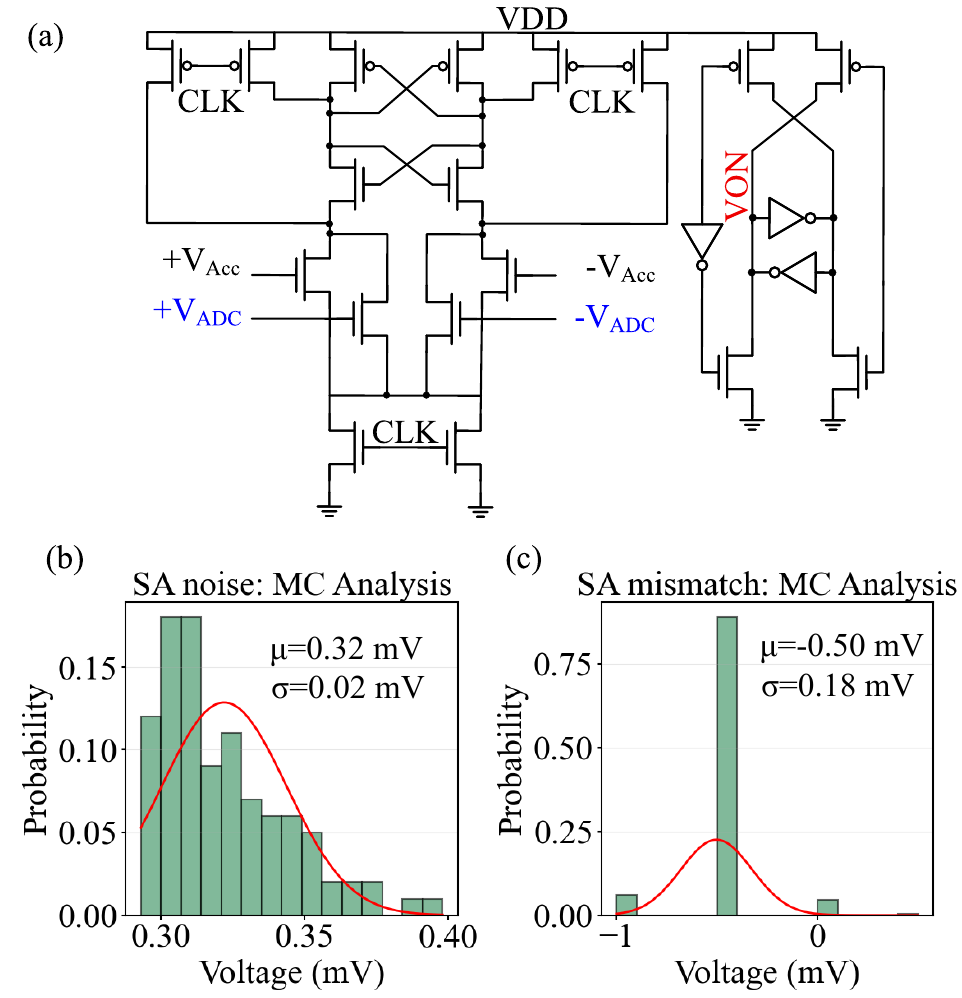}
  \caption{(a) SA schematic. (b) Monte Carlo simulation for SA noise. (c) Monte Carlo simulation for SA mismatch.}
  \label{SA_SCH_NOISE}
\end{figure}

\textbf{(3) Mismatch and noise analysis of SA.}
Fig. \ref{SA_SCH_NOISE} (a) shows the schematic of SA. We also performed a noise and mismatch MC simulation of SA using the same method as the buffer.  Simulation results in Fig. \ref{SA_SCH_NOISE} (b) and (c) indicate a mean noise of 0.32 mV and a mismatch of -0.5 mV. Both of the two values are significantly smaller compared to the 4.8 mV (LSB of the IMADC), demonstrating negligible SA noise and mismatch impact on the quantization precision of IMADC.
\subsection{Quantization-aware training (QAT) and noise-resilient Training (NRT)}
\label{noise training}

\begin{algorithm}[!t]
\caption{Noise resilient training.}\label{alg:alg1}
\begin{algorithmic}
\STATE 
\STATE Input: $X_t$, $W_t$ and $\sigma$
\STATE Output: $W_{t+1}$
\STATE \textbf{loop}
\STATE \hspace{0.4cm} Feed forward propagation: 
\STATE \hspace{0.5cm}$ Y_\mathbf{t} \gets  Q(W_t \cdot X_t) $
\STATE \hspace{0.5cm}$ Z_\mathbf{t} \gets  f(Y_t +\sigma) $
\STATE \hspace{0.4cm} Back-propagation propagation: 
\STATE \hspace{0.5cm}$ g_\mathbf{t} \gets$  calculate gradient based on $W_t$ and $f(W_t \cdot X_t)$
\STATE \hspace{0.5cm}$ W_\mathbf{t+1} \gets W_\mathbf{t}+\eta \times g_\mathbf{t} $
\STATE \textbf{end loop}
\end{algorithmic}
\end{algorithm}
We adopt QAT to enable low-bit weight quantization~\cite{li2016ternary}.
Let $m$ denote the per-layer (per-tensor) mean absolute weight:
\begin{equation}
m=\frac{1}{|\mathcal{W}|} \sum_{i \in \mathcal{W}}\left|W_i\right|,
\end{equation}
where $\mathcal W$ represents the number of all weights in the layer.

For ternary weight quantization (±1, 0): $\alpha = 0.7\, m.$
Weights are quantized as follows:

\begin{equation}
W_i= \begin{cases}+1, & W_i>\alpha \\ 0, & -\alpha \leq W_i \leq \alpha \\ -1, & W_i<-\alpha\end{cases}
\end{equation}

For signed 3-bit weight quantization (±3, ±2, ±1, 0):
$\alpha = 0.5\, m,
\beta = 1.5\, m, 
\gamma = 2.5\, m,$
Then weights are quantize as
\begin{equation}
W_i= \begin{cases}+3, & W_i>\gamma, \\ +2, & \beta<W_i \leq \gamma, \\ +1, & \alpha<W_i \leq \beta, \\ 0, & -\alpha \leq W_i \leq \alpha, \\ -1, & -\beta \leq W_i<-\alpha, \\ -2, & -\gamma \leq W_i<-\beta, \\ -3, & W_i<-\gamma .\end{cases}
\end{equation}

The thresholds $(\alpha, \beta, \gamma)$ are dynamically computed from the current full-precision weights at each training step. This on-the-fly calibration ensures balanced dynamic range coverage and encourages sparsity. The induced zeros allow for zero-skipping (ZOSKP) to reduce compute energy, as detailed in Section~\ref{sec:sw result}.

\begin{figure*}[t]
  \centering
  \includegraphics[width=1\linewidth]{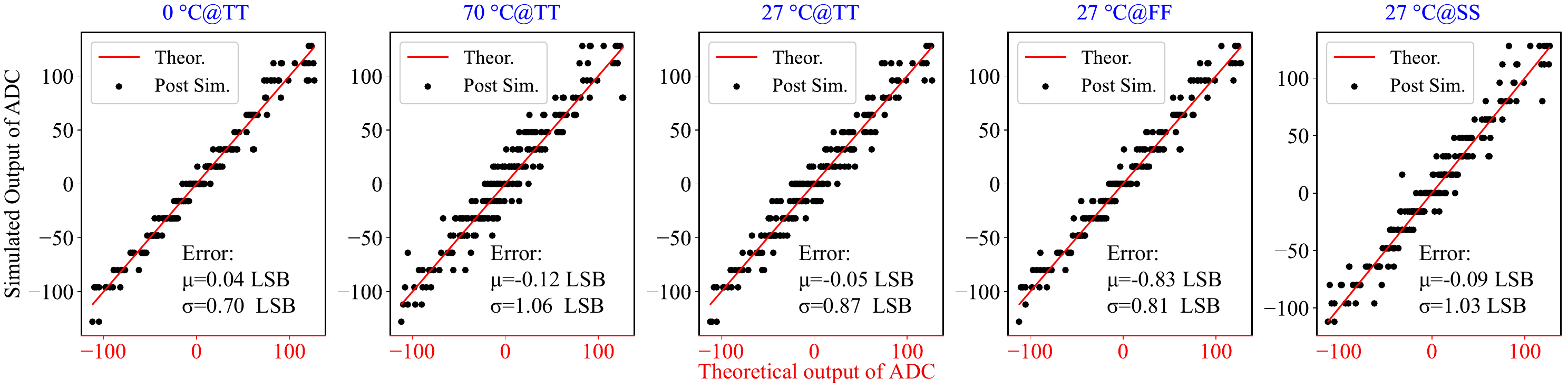}
  \caption{Post layout simulation versus theoretical 4-bit IMADC output across temperature variations (0°C, 70°C, 27°C) and different process corners (TT, FF, SS) (The step size of IMADC is 16, which is determined based on the range of the MAC).}
  \label{ADCINPUT_OUTPUT}
\end{figure*}

To address the detrimental impact of circuit non-idealities in the proposed architecture—including noise, mismatch and discharging currents of the bitcells—we employ NRT. This methodology compensates for inference accuracy degradation by explicitly modeling hardware imperfections during neural network training, drawing inspiration from QAT. As formalized in Algorithm \ref{alg:alg1}, NRT incorporates stochastic noise perturbations ($\delta$) sampled from empirical distributions derived from SPICE-level circuit simulations (Fig. \ref{ADCINPUT_OUTPUT}). These distributions capture the statistical signatures of hardware non-idealities, ensuring realistic noise profiles during training. 

In the feed-forward phase, the quantized MAC output $Y_t$ is corrupted by additive noise $\delta$, yielding a non-ideal activation output $Z_t$. This perturbed value propagates from first layer to last layer to emulate inference behavior in hardware. During backpropagation, however, gradient ($g_t$) is computed using the ideal output $f(W_t \cdot X_t)$ rather than $Z_t$. This decoupling ensures accurate gradient estimation by avoiding noise-induced bias in weight updates, while the feed-forward corruption forces the network to learn noise-invariant representations. The weight update rule consequently minimizes the divergence between ideal and noise-perturbed outputs across iterations, enhancing hardware robustness.


\begin{figure}[t]
  \centering
  \includegraphics[width=1\linewidth]{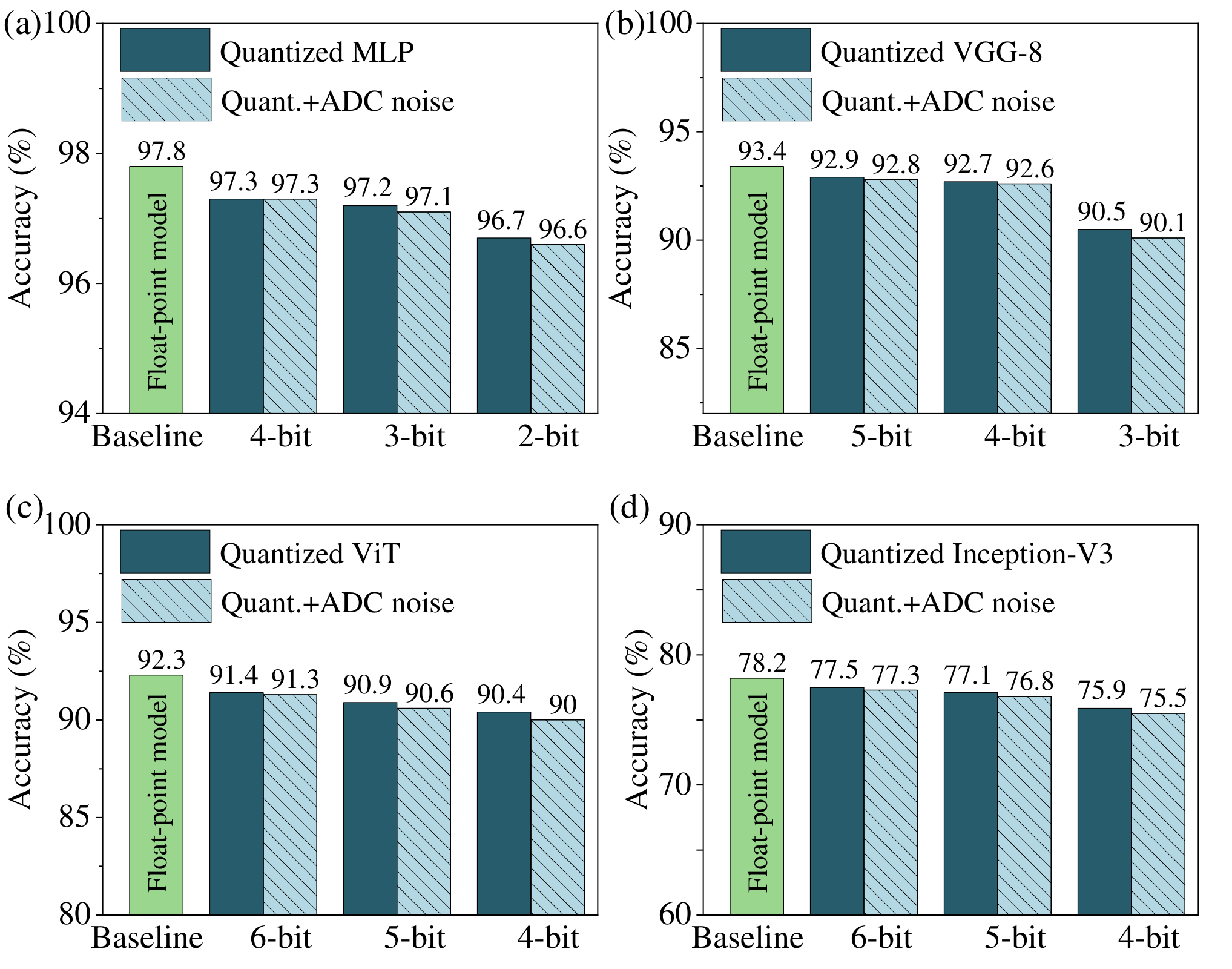}
  \caption{The inference accuracy results with hardware-simulated ADC noise and quantization model are compared with the software baselines using the floating-point model, as well as quantization model under different bit weight and ADC. (a) MLP with 2-bit weight, (b) VGG-8 with 2-bit weight, (c) ViT with 3-bit weight, and (d) Inception-V3 with 4-bit weight.} 
  \label{noise resilient training}
\end{figure}
\section{Simulation Result }
\label{sec:Result}
\subsection{Software results} 
\label{sec:sw result}
We choose three networks to evaluate the proposed method: one MLP network (784-128-128-10) on MNIST dataset, VGG-8 on CIFAR-10 dataset, and Vision Transformer (ViT) \cite{velickovic2017graph} on CIFAR-100 dataset and Inception-V3 on Tiny ImageNet.  

\begin{figure}[t]
  \centering
  \includegraphics[width=0.8\linewidth]{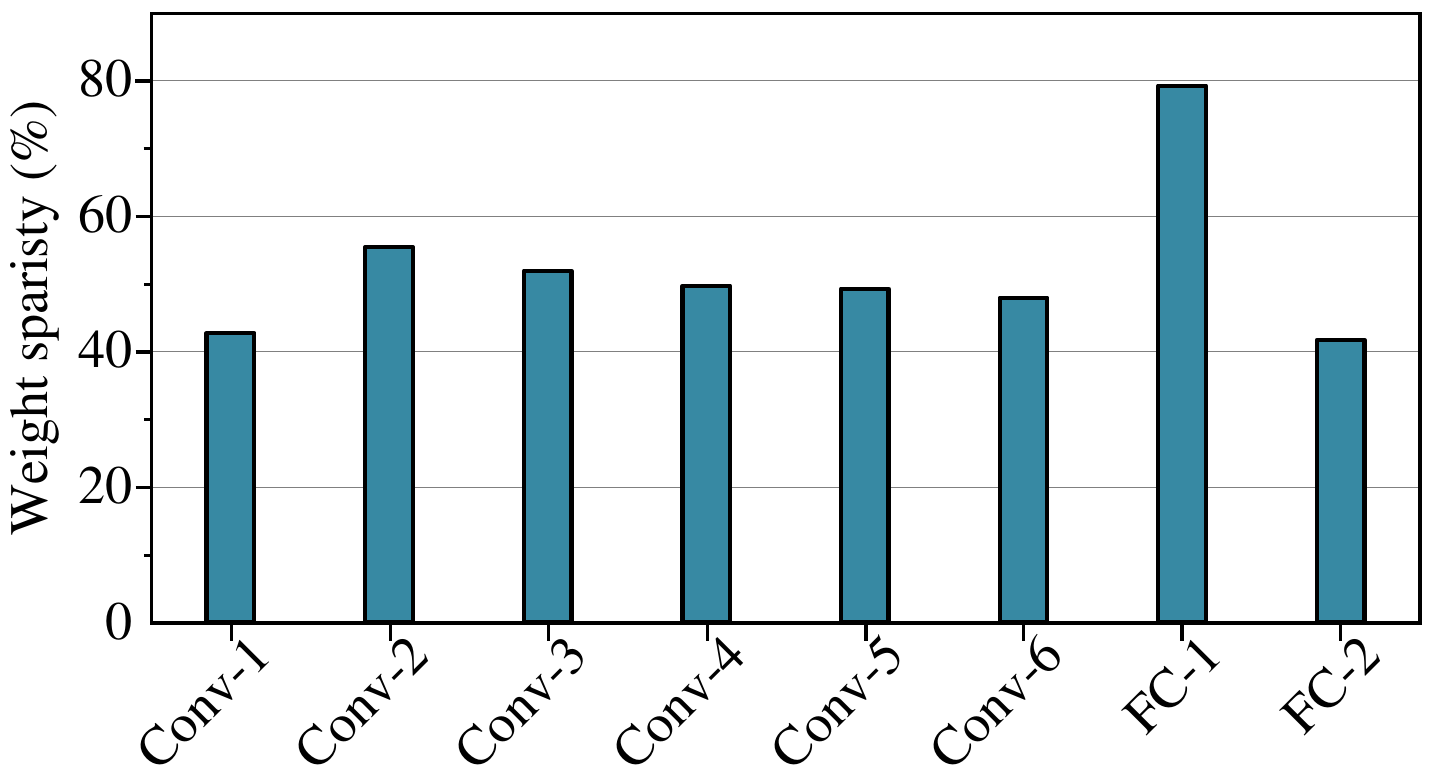}
  \caption{Weight sparsity result across each layer for VGG-8 on CIFAR-10 dataset after quantization of 2-bit weight.}
  \label{sparsity}
\end{figure}

By employing QAT, the four models achieve accuracies of 97.3\%, 92.9\%,  91.4\%, and 77.5\% at their highest resolutions, 4/2/4b for MLP, 5/2/5b for VGG-8, 6/3/6b for ViT, and 6/4/6b for Inception-V3 as shown in Fig. \ref{noise resilient training}.  Fig. \ref{sparsity} illustrates the weight sparsity after 2-bit quantization in the VGG-8 model. Across the eight layers of the quantized model, at least 40\% of the weights in the macro are zeros in every layer. Since the proposed dual 8T bitcell supports ZOSKP for weights, these zero-valued weights do not form discharge paths within the bitcell. This mechanism reduces the computational energy consumption in RBL by at least 40\%.

To conduct NRT, the proposed architecture is evaluated by using SPICE post layout simulation with parasitic parameters from layout in 65 nm CMOS technology. In Fig. \ref{ADCINPUT_OUTPUT}, the simulated 4-bit  IMADC outputs are compared against theoretical IMADC outputs across varying temperatures ($0^\circ \text{C}$, $27^\circ \text{C}$, and $70^\circ \text{C}$) and process corners (TT, FF, SS). Under worst-case temperature conditions (70°C), the standard deviation of the IMADC error increases to 1.31× that at room temperature (27°C), and under the SS process corner, it reaches 1.13× the value under the TT corner. These results demonstrate the robustness of the IMADC against both temperature and process variations because of the replica biasing.

Then the nominal error distribution \textit{N}(-0.05 LSB, 0.87 LSB) (27°C@TT in Fig. \ref{ADCINPUT_OUTPUT}) is used to inject noise into the four networks.  1) the MLP has less than 0.1\% accuracy decrease for 2-4 bit ADC compared to the quantized models due to the usage of NRT and shallow architecture of the model, which prevents deep accumulation of noise. 2) VGG-8 is a challenging model with deeper architecture. However, NRT can still restore precision to a comparable level of accuracy, with a decrease of less than 0.4\% compared to 3-5 bit quantized model. When this maximum error distribution \textit{N}(-0.12 LSB, 1.06 LSB) (70°C@TT in Fig. \ref{ADCINPUT_OUTPUT}) is added to the VGG-8 network, the inference accuracy drops is less than 0.55\% compared to 3-5 bit quantized model after re-training with NRT. This performance degradation remains minimal when compared to the 0.4\% accuracy drop observed at 27°C, highlighting the effectiveness of our approach in maintaining computational stability across thermal conditions. 3) ViT is a classical transformer architecture used in the image domain. Due to the higher computational complexity compared with MLP and VGG-8, it requires a higher $n_o$ for good performance. NRT demonstrates an accuracy degradation of less than 0.4\% when using 4–6 bit ADCs, compared to the quantized model.  4) Inception‑V3 is a very deep neural network that similarly requires a larger $n_o$. When using 4–6‑bit ADCs, the accuracy degradation remains below 0.5\% compared with the quantized model. The highest ADC bit resolution $n_o$ used for these four models results in $0.5\%$, $0.6\%$, $1.0\%$ and $0.9\%$ drop in accuracy from their floating point baselines (97.8\%, 93.4\%, 92.3\% and 78.2\%),  making them comparable to prior CIM deployments under practical non-idealities \cite{kim65nm8TSRAMIMC,jiang202240nm} and benchmarks\cite{dosovitskiy2020image, szegedy2016rethinking}.

\subsection{Hardware results}
\begin{figure}[t]
  \centering
  \includegraphics[width=0.95\linewidth]{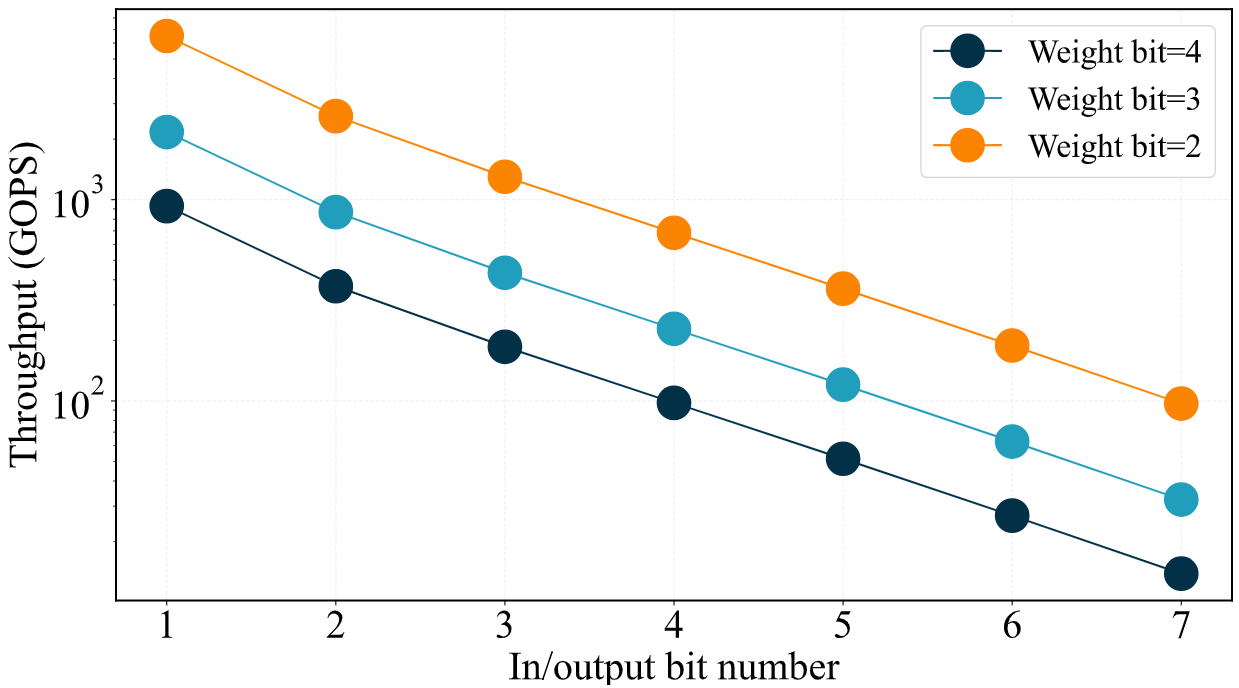}
  \caption{Throughput under different input, output, and weight resolutions.}
  \label{thoughtput}
\end{figure}

\textbf{Latency and throughput:} We assessed the latency of the system across three modes: proposed, PWM, and conventional BS, with respective latencies of \textit{n+2\textsuperscript{n}}, \textit{2\textsuperscript{n+1}}, and \textit{n2\textsuperscript{n}}, assuming the utilization of ramp ADC output and $n_i=n_o=n$. Fig. \ref{latency_and_overhead}(b) illustrates that our work yields 1.9× and 6.6× improvement of throughput over PWM mode and BS mode under 7-bit input/output resolution.  The curves of throughput versus input and output resolutions for 2-bit, 3-bit, and 4-bit weights are illustrated in Fig. \ref{thoughtput}. The throughput changes from  14 GOPS(7/4/7) to 6502 GOPS(1/2/1). We compared the throughput under high-precision configurations (4/4/4) with reference \cite{lee202328} (4/8/4). Our design achieves a throughput of 98 GOPS, which outperforms the 91 GOPS reported in the reference work.

\begin{figure*}[t]
  \centering
  \includegraphics[width=1\linewidth]{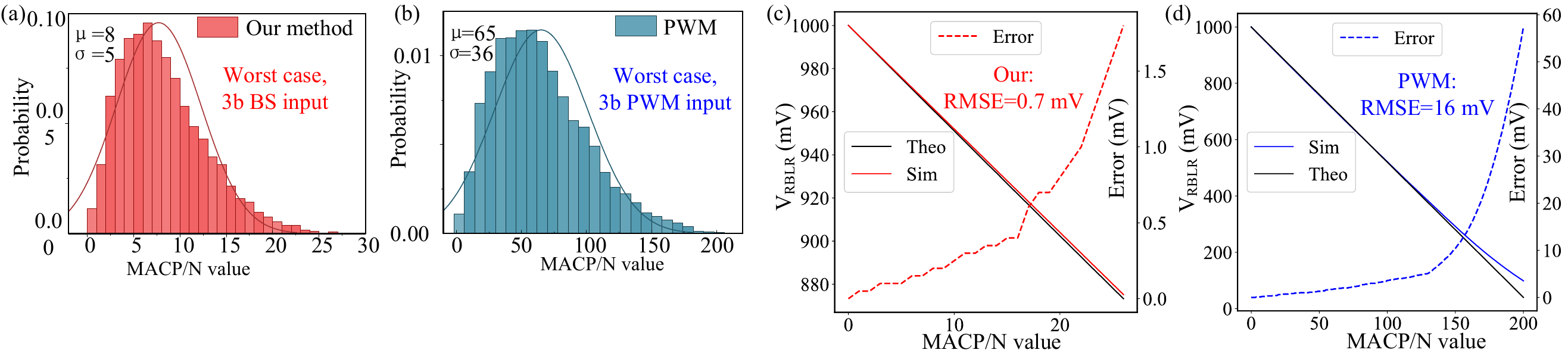}
  \caption{The partial MACP/N distribution of (a) Proposed method and (b) PWM mode based on 3-bit input for MNIST. Their corresponding voltages on RBL of (c) Proposed method and (d) PWM mode.}
  \label{VRBL}
\end{figure*}

\begin{figure}[t]
  \centering
  \includegraphics[width=1\linewidth]{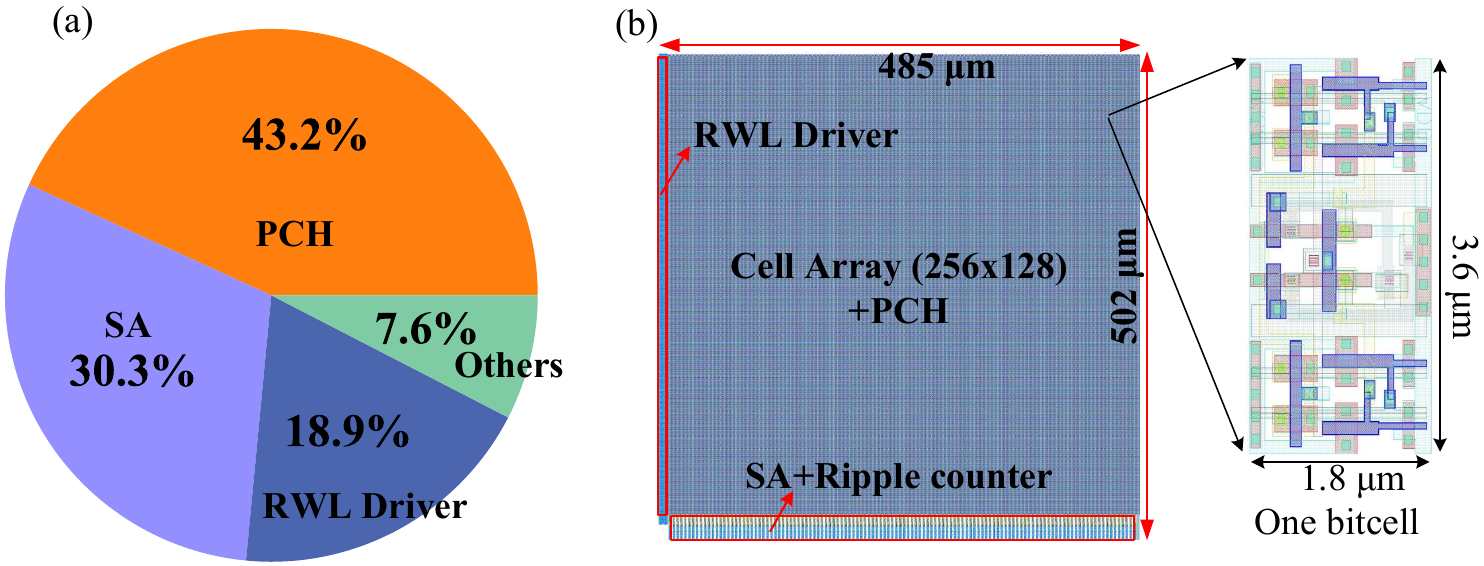}
  \caption{(a) Energy breakdown (4-bit input/output and 2-bit weight) based on post layout simulation. (b) Layouts of the macro and bitcell.}
  \label{enrgy and layout}
\end{figure}
\textbf{Enhanced linearity of MAC:}
During on-chip inference, the MAC ($MACP-MACN$) in every column is computed separately as MACP ($\operatorname{MACP}=\sum_{k=1}^N w_k x_k \quad\left(w_k \geq 0\right)$) and MACN ($\operatorname{MACN}=\sum_{k=1}^N w_k x_k \quad\left(w_k < 0\right)$), and subsequently input to SA as a differential signal. To determine the maximum voltage drop across RBL in on-chip inference, the distribution of MACP and MACN across all mini-batches in the test set for MNIST is used for the proposed and PWM method for $n_i=3$). In  Fig. \ref{VRBL} (a) and (b), the distribution of MACP in the worst-case scenario is depicted, while Fig. \ref{VRBL} (c) and (d) illustrates the corresponding RBL simulated voltage values, the theoretical values, and the error between them. Notably, the MACP/N distribution range in the PWM method is 7× greater than that in the our method reducing signal margin\cite{jhang2021challenges}. Furthermore, our method demonstrates a RBL voltage linearity that is $\approx$ 23× greater than that achieved by the PWM mode, as quantified by the Root Mean Squared Error (RMSE).

\begin{figure}[t]
  \centering
  \includegraphics[width=0.9\linewidth]{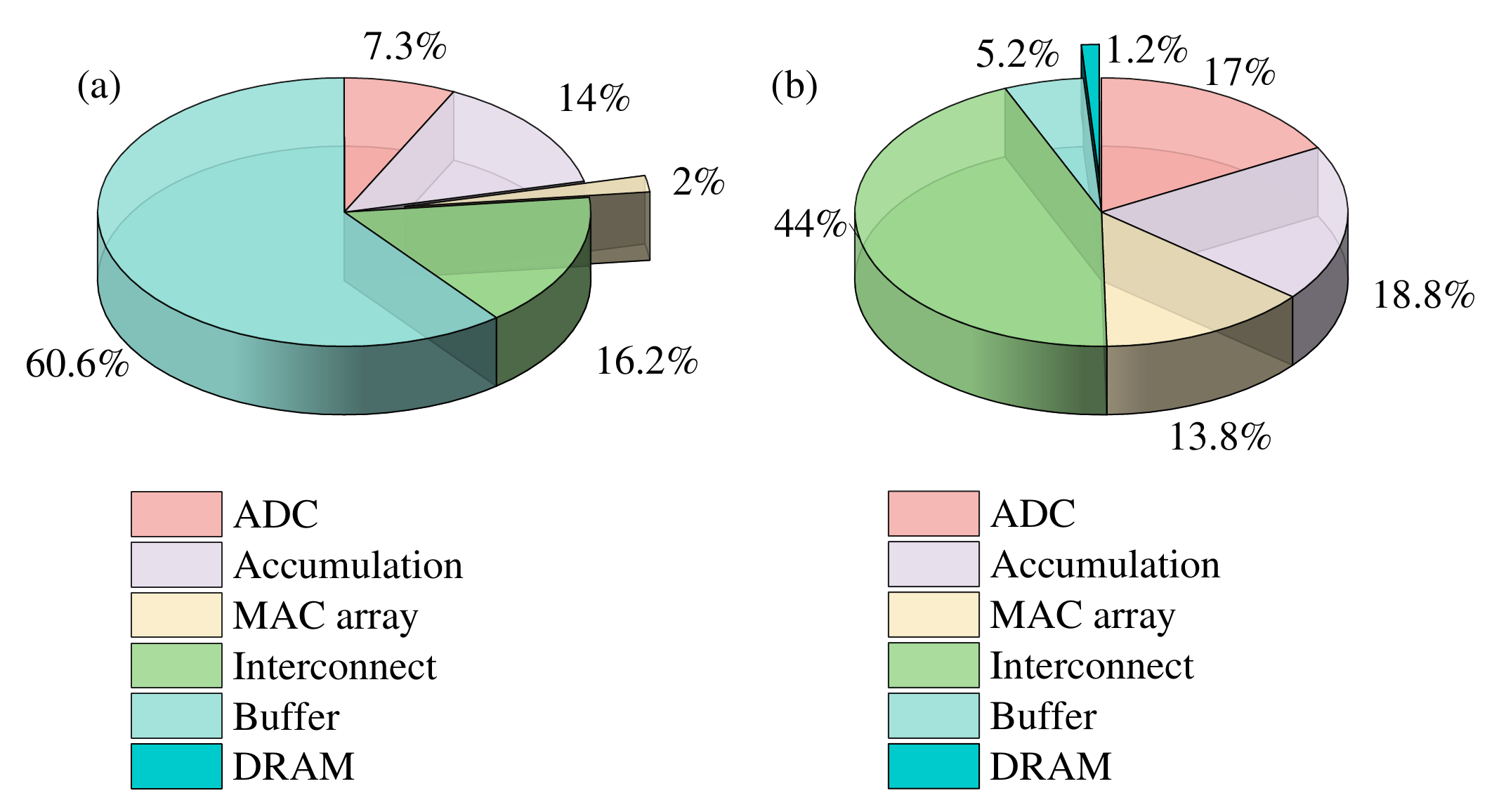}
  \caption{System performance for VGG-8 on CIFAR-10 dataset:  (a) latency breakdown and  (b) energy breakdown.}
  \label{system breakdown}
\end{figure}
\begin{figure}[!t]
  \centering
  \includegraphics[width=\linewidth]{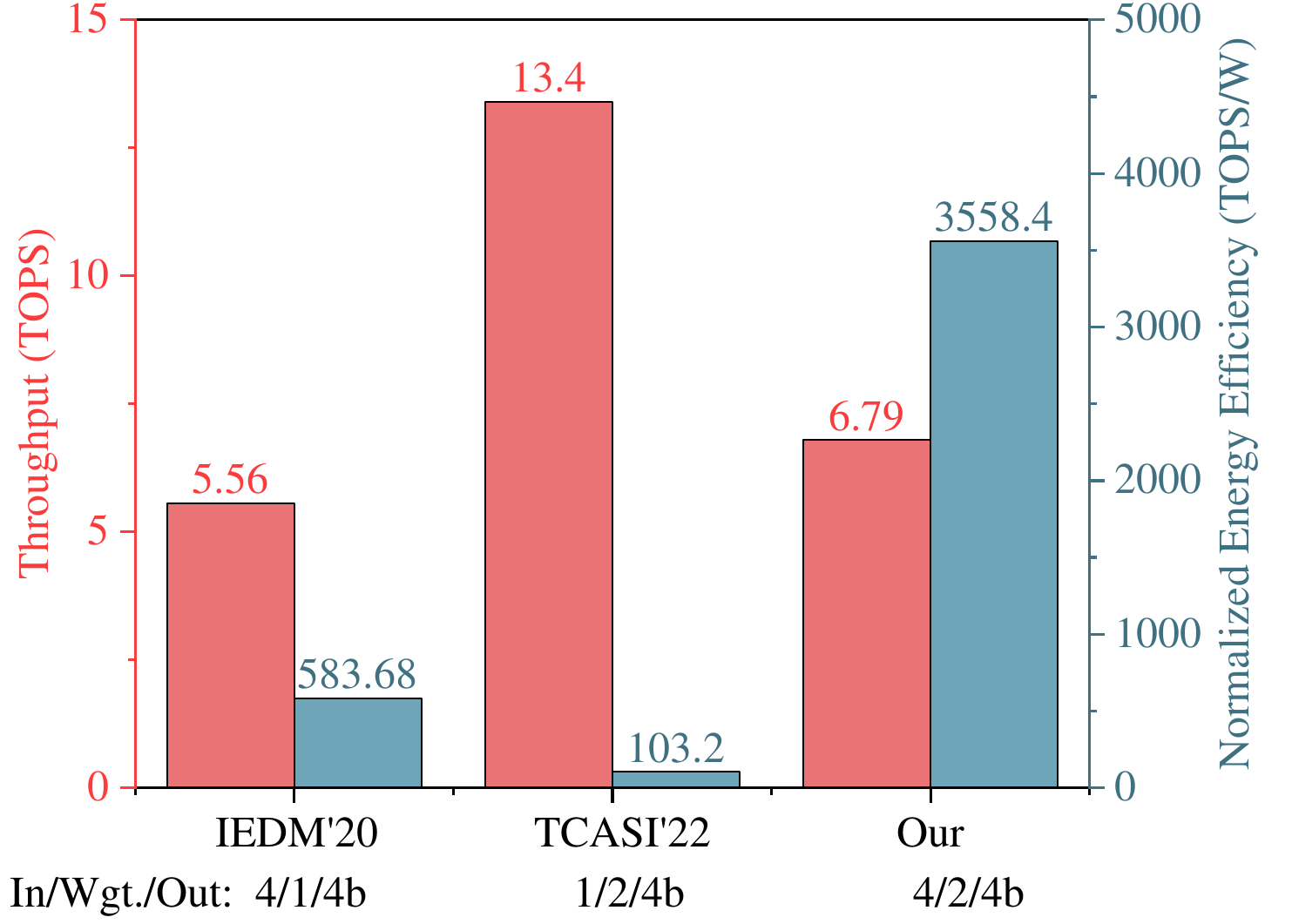}
  \caption{System throughput and energy efficiency comparison with previous works: IEDM'20\cite{peng2020benchmarking} and TCASI'22\cite{jiang2022enna}. Both of them are estimated by NeuroSim.}
  \label{system comparison}
\end{figure}

\begin{table*}[t]
\begin{center}
\begin{threeparttable}
\caption{Comparison with state-of-the-arts (ACIM, macro level)}
\begin{tabular}{|l|c|c|c|c|c|c|}
\hline
                            &   \begin{tabular}[r]{@{}l@{}} VLSI’25\cite{fu2025neuc}   \\  (Measurement) \end{tabular}               &  \begin{tabular}[r]{@{}l@{}} ISSCC’24\cite{wang202434}   \\  (Measurement) \end{tabular}        & \begin{tabular}[r]{@{}l@{}} JSSC’23\cite{lee202328}   \\  (Measurement) \end{tabular}    & \begin{tabular}[r]{@{}l@{}} TCASI’24\cite{yu2024dual}   \\  (Measurement) \end{tabular}       & \begin{tabular}[r]{@{}l@{}} JSSC’23\cite{yin2023cramming}   \\  (Measurement) \end{tabular}    & \textbf{\begin{tabular}[r]{@{}l@{}} This work   \\  (Post-layout simulation) \end{tabular}}      \\ \hline
Technology   (nm)          & 130                      & 14             & 28        & 65             & 65         & \textbf{65}             \\ \hline
Supply (V)                  & 0.9-1.2         & 0.8        & 1.2   & 0.45/0.8       & 0.7-1.2    & \textbf{0.5/0.8/1.0}    \\ \hline
Frequency   (MHz)           & 2.5                  & 200         & 100       & 200            & 50-210    & \textbf{200}            \\ \hline
Bitcell/Norm. area ($\mu m^2$)*                     & 8T/--                      & 5T-LF/8.4            & P-8T/4.3      & Dual 7T/5.76       & 1T ROM/-- & \textbf{Dual 8T/6.5}        \\ \hline
Computing   method       & Charge              & Charge        & Charge    & Current        & Charge     & \textbf{Current+Charge} \\ \hline
Array size                  &256×256                 & 64×32        & 256×64    & 528×128        & 64×64   & \textbf{256×128}        \\ \hline
Input/Output bit \#         & 1/1           &(1/2/8)/1-8       & 4/4       & 1/1-6          & 1-8/5    & \textbf{1-7/1-7}        \\ \hline
Weight bit \#                 & 4                        & 5/8              & 8         & 1.6/2.8/3.9    & 2/4/6/8        & \textbf{2-4}            \\ \hline
ADC type                    & --                   &SAR & Flash     & Ramp   (IMADC) & SAR        & \textbf{Ramp   (IMADC)} \\ \hline
ADC overhead**                & --                 & 4.7\%          & 84\%      & 27\%           & 13\%       & \textbf{3\%}            \\ \hline
Neuro \# per ADC            & --                        & 1              & 1         & 1              & 1         & \textbf{1}              \\ \hline
Throughput (GOPS)            & --                & 153.6            & 91     & 300-3900       & 13-54.6    & \textbf{14(7/4/7)}-6502(1/2/1)       \\ \hline
Area   efficiency (TOP/mm\textsuperscript{2})  &  --                 & 9.6          & 2.86 & 0.9-10.4       & 0.001-0.005  & \textbf{0.1(7/4/7)}-27(1/2/1)         \\ \hline
 \begin{tabular}[c]{@{}l@{}} Normalized area efficiency   \\  (TOP/mm\textsuperscript{2})***  \end{tabular}  & --            & 0.45     & 1.06 & 0.9-10.4        & 0.001-0.005    & \textbf{0.1(7/4/7)-27(1/2/1)}     \\ \hline
Energy   efficiency (TOP/W) & 143           & 22.64(1/8/8)     & 12.2 & 23.9-259       & 1.24-4.33    & \textbf{8.4(7/4/7)}-1023.2(1/2/1)     \\ \hline
 \begin{tabular}[c]{@{}l@{}} Normalized energy efficiency   \\  (TOP/W)**** \end{tabular}  & 1144            & 1448.9     & 1561.6 & 93.2-414.4       & 396.8-1385.6    & \textbf{1646.4-2046.4}     \\ \hline
\end{tabular}

    \begin{tablenotes}
    \item*Norm. area=area ×  (65 nm/technology)\textsuperscript{2}
      \item** ADC overhead=ADC area/MAC area 
      \item***Normalized area efficiency=area efficiency ×  (technology/65 nm)\textsuperscript{2}
      \item****Normalized energy efficiency=energy efficiency × input resolution × weight resolution × output resolution × (technology/65 nm)\cite{guo20195}
    \end{tablenotes}
  \end{threeparttable}
  \end{center}
\label{tab1}
\end{table*}

\textbf{Energy and Area efficiency (Macro level):}
Fig. \ref{enrgy and layout}(a) shows post simulated energy breakdown of our work when $n_i=4$ and weight is 2-bit. Pre-charging operations consume a considerable amount of energy (43.2\%). Notably, for a 4-bit input, the pre-charge procedure needs to be performed four times, significantly contributing to the total energy overhead. SAs consume 30.3\% of the energy and the energy consumed by the added voltage buffer is negligible, accounting for less than 2\%. The layout of core area is depicted in Fig. \ref{enrgy and layout}(b). The core size is 0.24 mm\textsuperscript{2} (Bitcell layout occupies 3.6 $\mu$m × 1.8 $\mu$m ) and 127 IMADCs only occupy only 3\% of the size of MAC array (1.5× improvement compared to 4.7\% in the best work \cite{wang202434}). 

Table I summarizes a performance comparison with the state-of-the-art ACIM. Under a 1/2/1b configuration, the macro achieves 1023.2 TOPS/W and 27 TOPS/mm\textsuperscript{2} while the macro maintains a performance of 8.4 TOPS/W and 0.1 TOPS/mm\textsuperscript{2} under the highest precision configuration of 7/4/7b. For a fair comparison, all energy efficiencies are normalized to 65 nm, 1/1/1 bit, according to the method in \cite{guo20195}, although different resolutions may carry different information content. Compared to the best prior art \cite{yin2023cramming},\cite{yu2024dual}, the macro offers 1.4× higher normalized energy efficiency and 2.6× greater normalized area efficiency, respectively. 
It is noteworthy that the proposed macro offers the highest level of reconfigurability in terms of precision for input (1-7b), weight (2-4b), and output (1-7b).   It is worth noting that the energy efficiency metrics of this work are derived from post-layout simulations, whereas other works in Table I report results based on silicon measurements. Although our post-layout simulations incorporate detailed parasitic extraction, a performance gap typically persists relative to real silicon measurements. This discrepancy primarily arises from manufacturing variations, packaging parasitic effects, and dynamic IR drop. To ensure a fair comparison, we have explicitly labeled the data type for each work in Table I.


\textbf{Energy and Area efficiency (System level):}
We conduct a comprehensive evaluation of the system-level overhead associated with running the VGG-8 model on the CIFAR-10 dataset. In this analysis, each component of the architecture is carefully modeled to ensure accuracy and fairness in comparison with prior work. The SRAM arrays as well as the ADCs are characterized directly from detailed SPICE simulations, which were previously presented to capture circuit-level delay/energy. For the remaining architectural modules—including the accumulation units, the on-chip interconnect fabric, input /output /global buffers, and the off-chip DRAM—we rely on cycle-accurate simulations using the NeuroSim framework \cite{peng2020dnn+}, which is widely adopted for benchmarking emerging in-memory computing systems. The system operates at a frequency of 200 MHz and a temperature of 300 K, using a 65 nm technology node with a wire width of 100 nm. The array size is 256$\times$128, and an H-tree interconnect is employed with a folding ratio of 4 in the layout. SRAM is used for the global, tile, and PE buffers. The ADC resolution is 4 bits. All other settings are kept unchanged. Note that the crossbar array–level performance originally evaluated using NeuroSim is replaced by SPICE-based macro-level results.

Fig. \ref{system breakdown} presents a detailed breakdown of latency and energy consumption across different architectural components. The results clearly indicate that system performance is largely dominated by the overhead of buffers and interconnects. This observation highlights a fundamental bottleneck in many in‑memory computing systems. Similar trends have been reported in other in‑memory computing architectures, such as ISAAC\cite{shafiee2016isaac}, PUMA\cite{ankit2019puma}, and FPSA\cite{ji2019fpsa}.


Overall, the evaluated design delivers a throughput of 6.79 TOPS and a normalized energy efficiency of 3558.4 TOPS/W  (defined as input bit × weight bit × output bit × energy efficiency) under a 4/2/4b configuration. As shown in Fig. \ref{system comparison}, compared with the baselines reported by Peng et al. \cite{peng2020benchmarking} (5.56 TOPS, normalized energy efficiency of 583.68 TOPS/W under a 4/1/4b configuration) and Jiang et al. \cite{jiang2022enna} (13.4 TOPS, normalized energy efficiency of 103.2 TOPS/W under a 1/2/4b configuration), our system achieves 6× improvement in normalized energy efficiency compared with the SOTA.

\section{Discussion}
We would also point out that while Flash ADCs have a faster conversion speed, one ADC has to be reused by many columns of the array (i.e. shared by many neurons) due to the tight area requirements of matching with the memory array. In fact, for the case of Flash ADC using the SA as comparator, one ADC has to be reused exactly $2^n$ times, making the overall throughput same as the case of the ramp ADC such as in \cite{li202240}. Moreover, our approach can tradeoff conversion time with ADC resolution flexibly. For Flash ADCs with separate physically instantiated comparators \cite{lee202328}, while the conversion time is fast, the area overhead is permanent and cannot be reconfigured flexibly. Lastly, we can also use coarse-fine approaches as done in \cite{lee202328} to check the sign of the input in 1 clock cycle and then start the ramp, we can reduce the clock cycles for n-bit ADC to $2^{n-1} + 1$. 

\section{Conclusion}
\label{sec:Conclusion}
This article addresses important issues of CIM:  multi-bit input, high linearity of MAC and large overhead of ADC. We enable multi-bit (1-7) input using BS with Charge-based accumulation of partial sums and multi-bit (1-7) output using a reconfigurable IMADC that reduces area overhead by shared references. The 256$\times$128 in-memory computing array can achieve high efficiency (1023.2 TOPS/W, 27 TOPS/mm\textsuperscript{2} at 1/2/1b for in/weight/out). Using NRT, its accuracy reductions are limited to only 0.1\%, 0.4\% and 0.4\% due to circuit non-idealities for the MLP model (2/2/2b) on MNIST, VGG-8 model (3/2/3b) on CIFAR-10 and ViT network (4/3/4b) on CIFAR-100 respectively. It increases throughput (1.9×) and linearity (23×) compared to PWM mode by BSCHA. Compared to conventional BS with digital accumulation after ADC, this method has 2.6×/1.4×/6.6× better normalized area-efficiency/normalized energy-efficiency/throughput respectively.  The system-level performance of VGG-8 on CIFAR-10 is evaluated using SPICE and NeuroSim, demonstrating a 6× enhancement in normalized energy efficiency.





\bibliography{IEEEref}

\begin{thebibliography}{10}
\providecommand{\url}[1]{#1}
\csname url@samestyle\endcsname
\providecommand{\newblock}{\relax}
\providecommand{\bibinfo}[2]{#2}
\providecommand{\BIBentrySTDinterwordspacing}{\spaceskip=0pt\relax}
\providecommand{\BIBentryALTinterwordstretchfactor}{4}
\providecommand{\BIBentryALTinterwordspacing}{\spaceskip=\fontdimen2\font plus
\BIBentryALTinterwordstretchfactor\fontdimen3\font minus \fontdimen4\font\relax}
\providecommand{\BIBforeignlanguage}[2]{{%
\expandafter\ifx\csname l@#1\endcsname\relax
\typeout{** WARNING: IEEEtran.bst: No hyphenation pattern has been}%
\typeout{** loaded for the language `#1'. Using the pattern for}%
\typeout{** the default language instead.}%
\else
\language=\csname l@#1\endcsname
\fi
#2}}
\providecommand{\BIBdecl}{\relax}
\BIBdecl

\bibitem{li2023towards}
G.~Li, J.~Ji, M.~Qin, W.~Niu, B.~Ren, F.~Afghah, L.~Guo, and X.~Ma, ``{Towards high-quality and efficient video super-resolution via spatial-temporal data overfitting},'' in \emph{2023 IEEE/CVF Conference on Computer Vision and Pattern Recognition (CVPR)}.\hskip 1em plus 0.5em minus 0.4em\relax IEEE, 2023, pp. 10\,259--10\,269.

\bibitem{hinton2012deep}
G.~Hinton, L.~Deng, D.~Yu, G.~E. Dahl, A.-r. Mohamed, N.~Jaitly, A.~Senior, V.~Vanhoucke, P.~Nguyen, T.~N. Sainath \emph{et~al.}, ``{Deep neural networks for acoustic modeling in speech recognition: The shared views of four research groups},'' \emph{IEEE Signal processing magazine}, vol.~29, no.~6, pp. 82--97, 2012.

\bibitem{zhang2022quantifying}
Q.~Zhang, X.~Cheng, Y.~Chen, and Z.~Rao, ``{Quantifying the knowledge in a DNN to explain knowledge distillation for classification},'' \emph{IEEE Transactions on Pattern Analysis and Machine Intelligence}, vol.~45, no.~4, pp. 5099--5113, 2022.

\bibitem{guo202428}
A.~Guo, C.~Xi, F.~Dong, X.~Pu, D.~Li, J.~Zhang, X.~Dong, H.~Gao, Y.~Zhang, B.~Wang \emph{et~al.}, ``{A 28-nm 64-kb 31.6-TFLOPS/W digital-domain floating-point-computing-unit and double-bit 6T-SRAM computing-in-memory macro for floating-point CNNs},'' \emph{IEEE Journal of Solid-State Circuits}, vol.~59, no.~9, pp. 3032--3044, 2024.

\bibitem{lee202328}
K.~Lee, J.~Kim, and J.~Park, ``{A 28-nm 50.1-TOPS/W P-8T SRAM compute-in-memory macro design with BL charge-sharing-based in-SRAM DAC/ADC operations},'' \emph{IEEE Journal of Solid-State Circuits}, vol.~59, no.~6, pp. 1926--1937, 2023.

\bibitem{chen2019eyeriss}
Y.-H. Chen, T.-J. Yang, J.~Emer, and V.~Sze, ``{Eyeriss v2: A flexible accelerator for emerging deep neural networks on mobile devices},'' \emph{IEEE Journal on Emerging and Selected Topics in Circuits and Systems}, vol.~9, no.~2, pp. 292--308, 2019.

\bibitem{yang2025efficient}
J.~Yang, R.~Mao, M.~Jiang, Y.~Cheng, P.-S.~V. Sun, S.~Dong, G.~Pedretti, X.~Sheng, J.~Ignowski, H.~Li \emph{et~al.}, ``{Efficient nonlinear function approximation in analog resistive crossbars for recurrent neural networks},'' \emph{Nature Communications}, vol.~16, no.~1, p. 1136, 2025.

\bibitem{cheon20232941}
S.~Cheon, K.~Lee, and J.~Park, ``{A 2941-TOPS/W charge-domain 10T SRAM compute-in-memory for ternary neural network},'' \emph{IEEE Transactions on Circuits and Systems I: Regular Papers}, vol.~70, no.~5, pp. 2085--2097, 2023.

\bibitem{guo202434}
A.~Guo, X.~Chen, F.~Dong, J.~Chen, Z.~Yuan, X.~Hu, Y.~Zhang, J.~Zhang, Y.~Tang, Z.~Zhang \emph{et~al.}, ``{34.3 a 22nm 64kb lightning-like hybrid computing-in-memory macro with a compressed adder tree and analog-storage quantizers for transformer and cnns},'' in \emph{2024 IEEE International Solid-State Circuits Conference (ISSCC)}, vol.~67.\hskip 1em plus 0.5em minus 0.4em\relax IEEE, 2024, pp. 570--572.

\bibitem{jhang202422}
C.-J. Jhang, W.-S. Khwa, P.-C. Wu, A.~S. Lele, P.-S. Wu, C.-E. Ke, T.-C. Chiu, Y.-C. Hung, W.-T. Hsu, J.-M. Hsu \emph{et~al.}, ``{A 22 nm 10.03-237.99 TOPS/W time-digital-hybrid SRAM compute-in-memory AI accelerator for GNN edge device applications},'' \emph{IEEE Transactions on Circuits and Systems for Artificial Intelligence}, vol.~1, no.~1, pp. 15--25, 2024.

\bibitem{twins8T}
X.~Si, J.-J. Chen, Y.-N. Tu, W.-H. Huang, J.-H. Wang, Y.-C. Chiu, W.-C. Wei, S.-Y. Wu, X.~Sun, R.~Liu \emph{et~al.}, ``{A twin-8T SRAM computation-in-memory unit-macro for multibit CNN-based AI edge processors},'' \emph{IEEE Journal of Solid-State Circuits}, vol.~55, no.~1, pp. 189--202, 2019.

\bibitem{yang202533}
J.~Yang, X.~Luo, Y.~Ke, Z.~Wang, H.~Shang, S.~Dong, Z.~Fu, X.~Yang, H.~Liu, and A.~Basu, ``{A 33.6--136.2-TOPS/W Nonlinear Analog Computing-in-Memory Macro for Multi-Bit LSTM Accelerator in 65-nm CMOS},'' \emph{IEEE Journal of Solid-State Circuits}, 2025.

\bibitem{roy2020memory}
K.~Roy, I.~Chakraborty, M.~Ali, A.~Ankit, and A.~Agrawal, ``{In-memory computing in emerging memory technologies for machine learning: An overview},'' in \emph{2020 57th ACM/IEEE Design Automation Conference (DAC)}.\hskip 1em plus 0.5em minus 0.4em\relax IEEE, 2020, pp. 1--6.

\bibitem{kim2023overview}
S.~Kim and H.-J. Yoo, ``{An overview of computing-in-memory circuits with DRAM and NVM},'' \emph{IEEE Transactions on Circuits and Systems II: Express Briefs}, vol.~71, no.~3, pp. 1626--1631, 2023.

\bibitem{zhang2017memory}
J.~Zhang, Z.~Wang, and N.~Verma, ``{In-memory computation of a machine-learning classifier in a standard 6T SRAM array},'' \emph{IEEE Journal of Solid-State Circuits}, vol.~52, no.~4, pp. 915--924, 2017.

\bibitem{zhang2023macc}
B.~Zhang, J.~Saikia, J.~Meng, D.~Wang, S.~Kwon, S.~Myung, H.~Kim, S.~J. Kim, J.-S. Seo, and M.~Seok, ``Macc-sram: A multistep accumulation capacitor-coupling in-memory computing sram macro for deep convolutional neural networks,'' \emph{IEEE Journal of Solid-State Circuits}, vol.~59, no.~6, pp. 1938--1949, 2023.

\bibitem{chen202115}
Z.~Chen, X.~Chen, and J.~Gu, ``{a 65nm 3T dynamic analog RAM-based computing-in-memory macro and CNN accelerator with retention enhancement, adaptive analog sparsity and 44TOPS/W system energy efficiency},'' in \emph{2021 IEEE International Solid-State Circuits Conference (ISSCC)}, vol.~64.\hskip 1em plus 0.5em minus 0.4em\relax IEEE, 2021, pp. 240--242.

\bibitem{kim2022overview}
D.~Kim, C.~Yu, S.~Xie, Y.~Chen, J.-Y. Kim, B.~Kim, J.~P. Kulkarni, and T.~T.-H. Kim, ``{An overview of processing-in-memory circuits for artificial intelligence and machine learning},'' \emph{IEEE Journal on Emerging and Selected Topics in Circuits and Systems}, vol.~12, no.~2, pp. 338--353, 2022.

\bibitem{KIMTCASIvoltagemode}
J.~Mu, H.~Kim, and B.~Kim, ``{SRAM-based in-memory computing macro featuring voltage-mode accumulator and row-by-row ADC for processing neural networks},'' \emph{IEEE Transactions on Circuits and Systems I: Regular Papers}, vol.~69, no.~6, pp. 2412--2422, 2022.

\bibitem{valavi201964}
H.~Valavi, P.~J. Ramadge, E.~Nestler, and N.~Verma, ``{A 64-tile 2.4-Mb in-memory-computing CNN accelerator employing charge-domain compute},'' \emph{IEEE Journal of Solid-State Circuits}, vol.~54, no.~6, pp. 1789--1799, 2019.

\bibitem{kim65nm8TSRAMIMC}
C.~Yu, T.~Yoo, K.~T.~C. Chai, T.~T.-H. Kim, and B.~Kim, ``{A 65-nm 8T SRAM compute-in-memory macro with column ADCs for processing neural networks},'' \emph{IEEE Journal of Solid-State Circuits}, vol.~57, no.~11, pp. 3466--3476, 2022.

\bibitem{kim2023neuro}
S.~Kim, S.~Kim, S.~Um, S.~Kim, K.~Kim, and H.-J. Yoo, ``{Neuro-CIM: ADC-less neuromorphic computing-in-memory processor with operation gating/stopping and digital--analog networks},'' \emph{IEEE Journal of Solid-State Circuits}, vol.~58, no.~10, pp. 2931--2945, 2023.

\bibitem{yang2025high}
J.~Yang, S.~Dong, Z.~Fu, H.~Shang, and A.~Basu, ``{High Energy-efficiency and Low latency In-Memory Computing using Analog Accumulator and In-Memory ADC with shared References},'' in \emph{2025 62nd ACM/IEEE Design Automation Conference (DAC)}.\hskip 1em plus 0.5em minus 0.4em\relax IEEE, 2025, pp. 1--7.

\bibitem{kim20231}
H.~Kim, J.~Mu, C.~Yu, T.~T.-H. Kim, and B.~Kim, ``{A 1-16b reconfigurable 80Kb 7T SRAM-based digital near-memory computing macro for processing neural networks},'' \emph{IEEE Transactions on Circuits and Systems I: Regular Papers}, vol.~70, no.~4, pp. 1580--1590, 2023.

\bibitem{rajanna2021sram}
V.~K. Rajanna, S.~Taneja, and M.~Alioto, ``{SRAM with In-Memory Inference and 90\% Bitline Activity Reduction for Always-On Sensing with 109 TOPS/mm 2 and 749-1,459 TOPS/W in 28nm},'' in \emph{ESSCIRC 2021-IEEE 47th European Solid State Circuits Conference (ESSCIRC)}.\hskip 1em plus 0.5em minus 0.4em\relax IEEE, 2021, pp. 127--130.

\bibitem{saragada2024process}
P.~K. Saragada and B.~P. Das, ``{Process-Variation-Aware In-Memory Computation With Improved Linearity Using On-Chip Configurable Current-Steering Thermometric DAC},'' \emph{IEEE Transactions on Circuits and Systems I: Regular Papers}, 2024.

\bibitem{shaik2024impact}
J.~B. Shaik, X.~Guo, and S.~Singhal, ``{Impact of aging and process variability on SRAM-based in-memory computing architectures},'' \emph{IEEE Transactions on Circuits and Systems I: Regular Papers}, 2024.

\bibitem{yu2024dual}
C.~Yu, H.~Jiang, J.~Mu, K.~T.~C. Chai, T.~T.-H. Kim, and B.~Kim, ``{A Dual 7T SRAM-Based Zero-Skipping Compute-In-Memory Macro With 1-6b Binary Searching ADCs for Processing Quantized Neural Networks},'' \emph{IEEE Transactions on Circuits and Systems I: Regular Papers}, 2024.

\bibitem{wang202434}
L.~Wang, W.~Li, Z.~Zhou, H.~Gao, Z.~Li, W.~Ye, H.~Hu, J.~Liu, J.~Yue, J.~Yang \emph{et~al.}, ``{34.9 a flash-SRAM-ADC-fused plastic computing-in-memory macro for learning in neural networks in a standard 14nm FinFET process},'' in \emph{2024 IEEE International Solid-State Circuits Conference (ISSCC)}, vol.~67.\hskip 1em plus 0.5em minus 0.4em\relax IEEE, 2024, pp. 582--584.

\bibitem{dong2025topkima}
S.~Dong, J.~Yang, X.~Peng, H.~Shang, Y.~Ke, X.~Yang, H.~Liu, and A.~Basu, ``{Topkima-Former: Low-Energy, Low-Latency Inference for Transformers Using Top-k In-Memory ADC},'' \emph{IEEE Transactions on Circuits and Systems I: Regular Papers}, 2025.

\bibitem{yin2025hybrid}
G.~Yin, Y.~Chen, M.~Lee, X.~Du, Y.~Ke, W.~Tang, Z.~Chen, M.~Zhou, J.~Yue, H.~Yang \emph{et~al.}, ``{Hybrid SRAM/ROM Compute-in-Memory Architecture for High Task-Level Energy Efficiency in Transformer Models With 8928-kb/$mm^2$ Density in 28nm CMOS},'' \emph{IEEE Journal of Solid-State Circuits}, 2025.

\bibitem{yin2023cramming}
G.~Yin, Y.~Chen, M.~Zhou, W.~Tang, M.~Lee, Z.~Yang, T.~Liao, X.~Du, V.~Narayanan, H.~Yang \emph{et~al.}, ``{Cramming More Weight Data Onto Compute-in-Memory Macros for High Task-Level Energy Efficiency Using Custom ROM With 3984-kb/mm 2 Density in 65-nm CMOS},'' \emph{IEEE Journal of Solid-State Circuits}, vol.~59, no.~6, pp. 1912--1925, 2023.

\bibitem{wang2023charge}
H.~Wang, R.~Liu, R.~Dorrance, D.~Dasalukunte, D.~Lake, and B.~Carlton, ``{A charge domain SRAM compute-in-memory macro with C-2C ladder-based 8-bit MAC unit in 22-nm FinFET process for edge inference},'' \emph{IEEE Journal of Solid-State Circuits}, vol.~58, no.~4, pp. 1037--1050, 2023.

\bibitem{DACoverheadbig}
K.~Roy, I.~Chakraborty, M.~Ali, A.~Ankit, and A.~Agrawal, ``{In-memory computing in emerging memory technologies for machine learning: An overview},'' in \emph{2020 57th ACM/IEEE Design Automation Conference (DAC)}.\hskip 1em plus 0.5em minus 0.4em\relax IEEE, 2020, pp. 1--6.

\bibitem{DACSARADC}
K.~Lee, S.~Cheon, J.~Jo, W.~Choi, and J.~Park, ``{A charge-sharing based 8T SRAM In-Memory Computing for edge DNN acceleration},'' in \emph{2021 58th ACM/IEEE Design Automation Conference (DAC)}.\hskip 1em plus 0.5em minus 0.4em\relax IEEE, 2021, pp. 739--744.

\bibitem{jiang2020c3sram}
Z.~Jiang, S.~Yin, J.-S. Seo, and M.~Seok, ``{C3SRAM: An in-memory-computing SRAM macro based on robust capacitive coupling computing mechanism},'' \emph{IEEE Journal of Solid-State Circuits}, vol.~55, no.~7, pp. 1888--1897, 2020.

\bibitem{dong2026cadc}
S.~Dong, J.~Yang, Y.~Ke, H.~Shang, and A.~Basu, ``Cadc: Crossbar-aware dendritic convolution for efficient in-memory computing,'' in \emph{2026 31st Asia and South Pacific Design Automation Conference (ASP-DAC)}.\hskip 1em plus 0.5em minus 0.4em\relax IEEE, 2026, pp. 667--673.

\bibitem{dong202015}
Q.~Dong, M.~E. Sinangil, B.~Erbagci, D.~Sun, W.-S. Khwa, H.-J. Liao, Y.~Wang, and J.~Chang, ``{A 351 TOPS/W and 372.4 GOPS compute-in-memory SRAM macro in 7nm FinFET CMOS for machine-learning applications},'' in \emph{2020 IEEE International Solid-State Circuits Conference-(ISSCC)}.\hskip 1em plus 0.5em minus 0.4em\relax IEEE, 2020, pp. 242--244.

\bibitem{gonugondla201842pj}
S.~K. Gonugondla, M.~Kang, and N.~Shanbhag, ``{A 42 pJ/decision 3.12 TOPS/W robust in-memory machine learning classifier with on-chip training},'' in \emph{2018 IEEE International Solid-State Circuits Conference-(ISSCC)}.\hskip 1em plus 0.5em minus 0.4em\relax IEEE, 2018, pp. 490--492.

\bibitem{jhang2021challenges}
C.-J. Jhang, C.-X. Xue, J.-M. Hung, F.-C. Chang, and M.-F. Chang, ``{Challenges and trends of SRAM-based computing-in-memory for AI edge devices},'' \emph{IEEE Transactions on Circuits and Systems I: Regular Papers}, vol.~68, no.~5, pp. 1773--1786, 2021.

\bibitem{li2016ternary}
F.~Li, B.~Liu, X.~Wang, B.~Zhang, and J.~Yan, ``{Ternary weight networks},'' \emph{arXiv preprint arXiv:1605.04711}, 2016.

\bibitem{han2025mitigating}
L.~Han, P.~Huang, Y.~Wang, Z.~Zhou, H.~Yang, Y.~Chen, X.~Liu, and J.~Kang, ``{Mitigating methodology of hardware non-ideal characteristics for non-volatile memory based neural networks},'' \emph{Science China Information Sciences}, vol.~68, no.~2, p. 122403, 2025.

\bibitem{yin2020xnor}
S.~Yin, Z.~Jiang, J.-S. Seo, and M.~Seok, ``{XNOR-SRAM: In-memory computing SRAM macro for binary/ternary deep neural networks},'' \emph{IEEE Journal of Solid-State Circuits}, vol.~55, no.~6, pp. 1733--1743, 2020.

\bibitem{dong2022backpropagation}
S.~Dong, Y.~Chen, Z.~Fan, K.~Chen, M.~Qin, M.~Zeng, X.~Lu, G.~Zhou, X.~Gao, and J.-M. Liu, ``{A backpropagation with gradient accumulation algorithm capable of tolerating memristor non-idealities for training memristive neural networks},'' \emph{Neurocomputing}, vol. 494, pp. 89--103, 2022.

\bibitem{AllenHolberg}
{Allen, P. E and Holberg, D.}, \emph{CMOS Analog Circuit Design}.\hskip 1em plus 0.5em minus 0.4em\relax Oxford University Press, 2011.

\bibitem{marisa2017pseudo}
T.~Marisa, T.~Niederhauser, A.~Haeberlin, R.~A. Wildhaber, R.~Vogel, J.~Goette, and M.~Jacomet, ``{Pseudo asynchronous level crossing ADC for ECG signal acquisition},'' \emph{IEEE transactions on biomedical circuits and systems}, vol.~11, no.~2, pp. 267--278, 2017.

\bibitem{velickovic2017graph}
P.~Velickovic, G.~Cucurull, A.~Casanova, A.~Romero, P.~Lio, Y.~Bengio \emph{et~al.}, ``{Graph attention networks},'' \emph{stat}, vol. 1050, no.~20, pp. 10--48\,550, 2017.

\bibitem{jiang202240nm}
H.~Jiang, W.~Li, S.~Huang, and S.~Yu, ``{A 40nm analog-input ADC-free compute-in-memory RRAM macro with pulse-width modulation between sub-arrays},'' in \emph{2022 IEEE Symposium on VLSI Technology and Circuits (VLSI Technology and Circuits)}.\hskip 1em plus 0.5em minus 0.4em\relax IEEE, 2022, pp. 266--267.

\bibitem{dosovitskiy2020image}
A.~Dosovitskiy, L.~Beyer, A.~Kolesnikov, D.~Weissenborn, X.~Zhai, T.~Unterthiner, M.~Dehghani, M.~Minderer, G.~Heigold, S.~Gelly \emph{et~al.}, ``{An image is worth 16x16 words: Transformers for image recognition at scale},'' \emph{arXiv preprint arXiv:2010.11929}, 2020.

\bibitem{szegedy2016rethinking}
C.~Szegedy, V.~Vanhoucke, S.~Ioffe, J.~Shlens, and Z.~Wojna, ``{Rethinking the inception architecture for computer vision},'' in \emph{Proceedings of the IEEE conference on computer vision and pattern recognition}, 2016, pp. 2818--2826.

\bibitem{peng2020benchmarking}
X.~Peng, W.~Chakraborty, A.~Kaul, W.~Shim, M.~S. Bakir, S.~Datta, and S.~Yu, ``{Benchmarking monolithic 3D integration for compute-in-memory accelerators: overcoming ADC bottlenecks and maintaining scalability to 7nm or beyond},'' in \emph{2020 IEEE International Electron Devices Meeting (IEDM)}.\hskip 1em plus 0.5em minus 0.4em\relax IEEE, 2020, pp. 30--4.

\bibitem{jiang2022enna}
H.~Jiang, S.~Huang, W.~Li, and S.~Yu, ``{ENNA: An efficient neural network accelerator design based on ADC-free compute-in-memory subarrays},'' \emph{IEEE Transactions on Circuits and Systems I: Regular Papers}, vol.~70, no.~1, pp. 353--363, 2022.

\bibitem{fu2025neuc}
H.~Fu, H.~Zheng, Y.~Zhou, X.~Wen, Y.~Chen, H.~Ren, X.~Lin, Z.~Zong, L.~Wu, and B.~Cheng, ``{NeuC-CIM: A 1.3 pJ/SOP Neuromorphic Charge-Domain Compute-in-Memory Macro for Spiking Neural Network},'' in \emph{2025 Symposium on VLSI Technology and Circuits (VLSI Technology and Circuits)}.\hskip 1em plus 0.5em minus 0.4em\relax IEEE, 2025, pp. 1--3.

\bibitem{guo20195}
R.~Guo, Y.~Liu, S.~Zheng, S.-Y. Wu, P.~Ouyang, W.-S. Khwa, X.~Chen, J.-J. Chen, X.~Li, L.~Liu \emph{et~al.}, ``{A 5.1 pJ/neuron 127.3 us/inference RNN-based speech recognition processor using 16 computing-in-memory SRAM macros in 65nm CMOS},'' in \emph{2019 Symposium on VLSI Circuits}.\hskip 1em plus 0.5em minus 0.4em\relax IEEE, 2019, pp. C120--C121.

\bibitem{peng2020dnn+}
X.~Peng, S.~Huang, H.~Jiang, A.~Lu, and S.~Yu, ``{DNN+ NeuroSim V2. 0: An end-to-end benchmarking framework for compute-in-memory accelerators for on-chip training},'' \emph{IEEE Transactions on Computer-Aided Design of Integrated Circuits and Systems}, vol.~40, no.~11, pp. 2306--2319, 2020.

\bibitem{shafiee2016isaac}
A.~Shafiee, A.~Nag, N.~Muralimanohar, R.~Balasubramonian, J.~P. Strachan, M.~Hu, R.~S. Williams, and V.~Srikumar, ``{ISAAC: A convolutional neural network accelerator with in-situ analog arithmetic in crossbars},'' \emph{ACM SIGARCH Computer Architecture News}, vol.~44, no.~3, pp. 14--26, 2016.

\bibitem{ankit2019puma}
A.~Ankit, I.~E. Hajj, S.~R. Chalamalasetti, G.~Ndu, M.~Foltin, R.~S. Williams, P.~Faraboschi, W.-m.~W. Hwu, J.~P. Strachan, K.~Roy \emph{et~al.}, ``{PUMA: A programmable ultra-efficient memristor-based accelerator for machine learning inference},'' in \emph{Proceedings of the twenty-fourth international conference on architectural support for programming languages and operating systems}, 2019, pp. 715--731.

\bibitem{ji2019fpsa}
Y.~Ji, Y.~Zhang, X.~Xie, S.~Li, P.~Wang, X.~Hu, Y.~Zhang, and Y.~Xie, ``{FPSA: A full system stack solution for reconfigurable ReRAM-based NN accelerator architecture},'' in \emph{Proceedings of the Twenty-Fourth International Conference on Architectural Support for Programming Languages and Operating Systems}, 2019, pp. 733--747.

\bibitem{li202240}
W.~Li, X.~Sun, S.~Huang, H.~Jiang, and S.~Yu, ``{A 40-nm MLC-RRAM compute-in-memory macro with sparsity control, on-chip write-verify, and temperature-independent ADC references},'' \emph{IEEE Journal of Solid-State Circuits}, vol.~57, no.~9, pp. 2868--2877, 2022.

\end{thebibliography}
\bibliographystyle{IEEEtran}

\end{document}